\documentclass{aa}  
\usepackage{threeparttable}
\usepackage{natbib}
\usepackage{xcolor}
\usepackage{placeins}

\usepackage{microtype}
\usepackage{hyperref}

\usepackage{graphicx}
\usepackage{txfonts}

\begin{document}

   \title{The Fornax 3D project: Non-linear colour-metallicity relation of globular clusters\thanks{Based on observations collected at the ESO Paranal La Silla Observatory,
Chile, Prog. 296.B-5054(A)}}

\titlerunning{Fornax 3D: non-linear colour-metallicity relation of globular clusters}

   \author{K. Fahrion\inst{1}
          \and
          M. Lyubenova\inst{1}
          \and
          M. Hilker\inst{1}
          \and
           G. van de Ven\inst{2} 
           \and
          J. Falc\'{o}n-Barroso\inst{3}\fnmsep\inst{4}          
          \and
          R. Leaman\inst{5}
          \and
           I. Mart\'{i}n-Navarro\inst{6}\fnmsep\inst{4}\fnmsep\inst{7}\fnmsep\inst{5}
          \and
          A. Bittner\inst{1}
          \and
           L. Coccato\inst{1}
           \and
           E. M. Corsini\inst{8}\fnmsep\inst{9}         
           \and
           D. A. Gadotti\inst{1}
          \and
          E. Iodice\inst{1}\fnmsep\inst{10}
           \and           
          R. M. McDermid\inst{11}
          \and 
          F. Pinna\inst{5}
          \and
           M. Sarzi\inst{12}\fnmsep\inst{13}
           \and
           S. Viaene\inst{14}
          \and
          P. T. de Zeeuw\inst{15}\fnmsep\inst{16}
          \and
          L. Zhu\inst{17}}

   \institute{European Southern Observatory, Karl-Schwarzschild-Stra\ss{}e 2, 85748 Garching bei M\"unchen, Germany\\
   				\email{kfahrion@eso.org}
         \and 
             Department of Astrophysics, University of Vienna, T\"urkenschanzstrasse 17, 1180 Wien, Austria
             \and 
			Instituto de Astrof\'isica de Canarias, Calle Via L\'{a}ctea s/n, 38200 La Laguna, Tenerife, Spain
			\and 
			Depto. Astrof\'isica, Universidad de La Laguna, Calle Astrof\'isico Francisco S\'{a}nchez s/n, 38206 La Laguna, Tenerife, Spain
             \and 
             Max-Planck-Institut f\"ur Astronomie, K\"onigstuhl 17, 69117 Heidelberg, Germany
               \and 
               Instituto de Astrof\'{i}sica de Canarias, E-38200 La Laguna, Tenerife, Spain       
			\and 
               University of California Santa Cruz, 1156 High Street, Santa Cruz, CA 95064, USA
               \and 
                Dipartimento di Fisica e Astronomia 'G. Galilei', Universit\`a di Padova,  vicolo dell'Osservatorio 3, I-35122 Padova, Italy
			\and 
			INAF--Osservatorio Astronomico di Padova, vicolo dell'Osservatorio 5, I-35122 Padova, Italy
			 \and 
             INAF-Astronomical Observatory of Capodimonte, via Moiariello 16, I-80131, Napoli, Italy
               \and 
             Department of Physics and Astronomy, Macquarie University, North Ryde, NSW 2109, Australia
             \and 
             Armagh Observatory and Planetarium, College Hill, Armagh, BT61 9DG, Northern Ireland, UK
             \and 
             Centre for Astrophysics Research, University of Hertfordshire, College Lane, Hatfield AL10 9AB, UK
             \and 
             Sterrenkundig Observatorium, Universiteit Gent, Krijgslaan 281, B-9000, Gent, Belgium
             \and 
             Sterrewacht Leiden, Leiden University, Postbus 9513, 2300 RA Leiden, The Netherlands
             \and 
             Max-Planck-Institut f\"ur extraterrestrische Physik, Gie\ss{}enbachstraße 1, 85748 Garching bei M\"unchen, Germany
             \and 
             Shanghai Astronomical Observatory, Chinese Academy of Sciences, 80 Nandan Road, Shanghai 200030, China
             }

   \date{}

 
  \abstract
  {Globular cluster (GC) systems of massive galaxies often show a bimodal colour distribution. This has been interpreted as a metallicity bimodality, created by a two-stage galaxy formation where the red, metal-rich GCs were formed in the parent halo and the blue metal-poor GCs were accreted. This interpretation, however, crucially depends on the assumption that GCs are exclusively old stellar systems with a linear colour-metallicity relation (CZR). The shape of the CZR and range of GC ages are currently under debate, because their study requires high quality spectra to derive reliable
  stellar population properties. We determined metallicities with full spectral fitting from a sample of 187 GCs with high spectral signal-to-noise ratio in 23 galaxies of the Fornax cluster that were observed as part of the Fornax 3D project. The derived CZR from this sample is non-linear and can be described by a piecewise linear function with a break point at ($g - z$) $\sim$ 1.1 mag. The less massive galaxies in our sample ($M_\ast < 10^{10} M_\sun$) appear to have slightly younger GCs, but the shape of the CZR is insensitive to the GC ages. Although the least massive galaxies lack red, metal-rich GCs, a non-linear CZR is found irrespective of the galaxy mass, even in the most massive galaxies ($M_\ast \geq 10^{11} M_\sun$).
 Our CZR predicts narrow unimodal GC metallicity distributions for low mass and broad unimodal distributions for very massive galaxies, dominated by a metal-poor and metal-rich peak, respectively, and bimodal distributions for galaxies with intermediate masses (10$^{10}$ $\leq$ $M_\ast < 10^{11} M_\sun$) as a consequence of the relative fraction of red and blue GCs. The diverse metallicity distributions challenge the simple differentiation of GC populations solely based on their colour.}
   \maketitle
%

\section{Introduction}
\label{sect:intro}
Cosmological simulations provide a framework of galaxy formation and evolution via the hierarchical mergers of smaller galaxies, but the assembly of individual galaxies is challenging to constrain observationally. Globular clusters (GCs) are traditionally used to study galaxy assembly due to their ubiquitous occurrence in all massive galaxies ($M_\ast > 10^{9} M_\sun$, see \citealt{Brodie2006}). Their potential as tracers of galaxy evolution is based on their old ages ($\gtrsim 10$ Gyr, \citealt{Puzia2005, Strader2005}), which sets their formation at a redshift of $z \gtrsim 2$, coinciding with the peak of cosmic star formation \citep{Madau2014, ElBadry2019, ReinaCampos2019}. The survival of GCs until today allows us to view them as fossil records that have the chemodynamical properties of their origin encapsulated in their stellar population properties and orbital parameters which change only slowly over time (e.g. \citealt{Brodie2006, Beasley2008, Harris2016}).

In context of galaxy assembly, the metallicity distribution function (MDF) of GCs is of particular importance. If GCs trace the metallicity of their birthplace, the diverse merger histories of major galaxies as predicted from cosmological simulations (e.g. \citealt{Kruijssen2019}) translate into diverse shapes of the MDF and consequently, the shape of the MDF can put constraints on the merger history. 

In many galaxies, the MDF was found to have a bimodal shape with a metal-poor ([Fe/H] $\sim -1.5$ dex) and a metal-rich component ([Fe/H] $\sim -0.5$ dex), for example in the Milky Way (MW, e.g. \citealt{Harris1979, Zinn1985}), Centaurus\,A (NGC\,5128, \citealt{Beasley2008}), and the Sombrero galaxy (M104, \citealt{AlvesBrito2011}). The bimodality of the MDF is often interpreted as direct result of two-stage formation of massive galaxies (e.g. \citealt{Zepf1993, Beasley2002, Brodie2006, Harris2010, Forbes2011, Cantiello2014, Kartha2016}): the metal-rich GCs are thought to have formed primarily in-situ in the parent halo, whereas the metal-poor GCs formed in less massive galaxies and were accreted during the assembly of the host (e.g. \citealt{Cote1998, Hilker1999, Cote2000, Katz2014}). 
However, in some galaxies such as M31, the bimodality of the GC MDF is debated (e.g. \citealt{Barmby2000, Galleti2009}), with recent studies indicating even a trimodal distribution \citep{Caldwell2016}. Broad multimodal MDFs were suggested in a photometric study of brightest cluster galaxies, the most massive early-type galaxies (ETGs, \citealt{Harris2014, Harris2016, Harris2017}), based on unimodal colour distributions, making the shape of the GC MDF a heavily discussed topic nowadays. 

Because a detailed study of extragalactic GC MDFs requires time-expensive spectroscopy of individual GCs, often optical photometric studies of GC systems are used to infer the MDF from a colour distribution. These studies have shown that most massive galaxies have bimodal GC colour distributions (e.g. \citealt{Kundu2001, Larsen2001, Peng2006}), and because GCs are usually old stellar systems (e.g. \citealt{Strader2005}), this colour bimodality is usually translated into a bimodal MDF. However, this convertion crucially depends on the shape of colour-metallicity relation (CZR, e.g. whether it is linear or not)\footnote{We chose CZR as abbreviation to prevent possible confusion with the term "colour-magnitude relation".}. Both \cite{Richtler2006} and \cite{Yoon2006} suggested that a strongly non-linear CZR can produce bimodal colour distributions from broad unimodal metallicity distributions, challenging the view of a simple two-phase galaxy formation. 

Also the choice of colour can affect the inferred colour distributions. Bimodal distributions are more commonly seen when using optical colours, while optical-near-infrared colours can show unimodal colour distributions for the same GC system (e.g. \citealt{Blakeslee2012, ChiesSantos2012, Cho2016}). These colour-colour non-linearities suggest a underlying non-linearity of the CZR in some colours, with the optical-near-infrared colours being least sensitive \citep{Cantiello2007}. However, the S0 galaxy NGC\,3115 was found to show both non-linearities in colour-colour space as well as a bimodal metallicity distribution \citep{Cantiello2014}.

Due to the lack of large homogeneous samples of spectroscopic GC metallicities, there is no consensus on the shape of the CZR. Using the few spectroscopic GC metallicities available at that time, \cite{Peng2006} presented a piecewise linear CZR with a breakpoint at ($g - z) \sim 1.0$ mag. A similar description was found by \cite{Usher2012}, however, with a breakpoint at bluer colours. Their result was based on a diverse sample of GC metallicities from five different massive galaxies, brought to the photometry scheme of the SAGES Legacy Unifying Globulars and Galaxies Survey (SLUGGS; \citealt{Brodie2014}). While these two studies combined GC metallicities from different galaxies, \cite{Sinnott2010} and \cite{Harris2017} proposed a CZR described by a quadratic function based on literature metallicities of Centaurus\,A. Recently, \cite{Villaume2019} presented a linear CZR based on metallicities of 177 GCs of M87, the central galaxy of the Virgo cluster.  

The different results on the shape of the CZR might be connected to different measurement techniques. But, it could also indicate that the CZR is not universal and possibly depends on the host galaxy or the environment. For example, \cite{Villaume2019} found a lack of metal-poor GCs in M87 compared to other systems such as the MW. \cite{Usher2015} found indications for a CZR that varies from galaxy to galaxy and suggested this might be caused by different GC age distributions because the CZR strongly depends on the assumption that GCs are old stellar systems. Although many spectroscopic studies of extragalactic GCs have found generally old ages of $\gtrsim$ 12 Gyr (e.g. \citealt{Cohen1998, Forbes2001, Puzia2005, Norris2008}), there are also examples of younger GCs in a few galaxies (e.g. \citealt{Chandar2006, Sharina2006, Hempel2007, Martocchia2018, Sesto2018, Usher2019}). These might cause deviations in the CZR due to the age-metallicity degeneracy \citep{Worthey1994}.

In this paper, we present a non-linear CZR that was obtained using a sample of 187 GCs of 23 galaxies that were observed as part of the Fornax 3D project (F3D, \citealt{F3D_Survey}), a magnitude-limited survey with the Multi Unit Spectroscopic Explorer (MUSE) on the Very Large Telescope of bright galaxies within the virial radius of the Fornax cluster. The GCs we use in this work are a sub-sample of the GC catalogue presented in \citealt[(hereafter paper I)]{Fahrion2020b}, in which we tested the ability of GCs as tracers of kinematics and stellar population properties. Because F3D covers both ETGs and late-type galaxies (LTGs) with masses ranging between $10^8$ and $10^{11} M_\sun$, we can explore a sample of GCs over a variety of galaxy masses. In paper I, we found that especially the red GCs closely trace the metallicity of the host galaxy, even in the inner parts of galaxies, while the blue GCs are significantly more metal-poor at all radii. In the current paper, we extend the analysis of GC metallicities to derive a CZR from a well sampled range of GC colours without the need to combine metallicity measurements from different studies. While most previous works have focused on rather massive galaxies with $M_\ast > 10^{10} M_\sun$, we can explore the effect of the host galaxy on the CZR because of the broad mass range of galaxies in F3D.

We describe the GC sample in the next section and the methods for the stellar population measurements are briefly described in Sect. \ref{sect:stellar_pop_methods}. Sect. \ref{sect:results} presents our results for the CZR and describes tests to validate the measured metallicities. We also present estimates of GC ages and the mass-metallicity relation (MZR) of our sample. In Sect. \ref{sect:discussion}, we discuss our findings in relation to the literature and describe possible implications for galaxy evolution. We summarise and present our results and conclusions in Sect. \ref{sect:conclusions}.

\section{Globular cluster sample}
\label{sect:sample}
We described the extraction and basic analysis of a sample of 722 spectroscopically confirmed GCs in 32 galaxies of the Fornax cluster in paper I. These galaxies were observed as part of F3D and details on the MUSE observations can be found in \cite{F3D_Survey} and \cite{Iodice2019}. The spectra of these GCs were directly extracted from the MUSE cubes and for each GC, a spectrum of the local galaxy background was subtracted. Because F3D targets the central regions of galaxies (up to $\sim$ 3 $R_\text{eff}$), this cleaning process is necessary to remove the galaxy contribution that otherwise heavily contaminates the GC spectrum. Each GC spectrum was then classified by its spectral signal-to-noise ratio (S/N). Compared to the ACSFCS catalogue \citep{Jordan2015}, we reached a completeness of $\sim$ 50 \% at an absolute $g$-band magnitude of \mbox{$M_g \sim -8$ mag}.

In the first paper, we derived line-of-sight (LOS) velocities from all GC spectra with S/N $\geq 3$ \AA$^{-1}$, and metallicities for the GCs with S/N $\geq 8$ \AA$^{-1}$. For the present work, we only include GCs with a galactocentric distance $r_\text{gal}$ $\geq$ 15\arcsec\,because testing has shown that the spectra of GCs with small galactocentric distances can still be contaminated by residual galaxy light that strongly varies in the central regions. These GCs can be biased to higher metallicities because the host galaxy tends to be more metal-rich than the GCs, especially in the centre. From the initial sample of 722 GCs, this cut in S/N and galactocentric distance leaves a sub-sample of 187 GCs in 23 galaxies. Table \ref{tab:sample} gives an overview of this sub-sample and lists the number of available GC metallicities per galaxy. The majority of GCs were found in ETG hosts.
\begin{table*}
\centering
\caption{Overview of the GC sample used in this work.}
\label{tab:sample}
\begin{tabular}{c c c c c c c c}\hline
Galaxy  & Altern. name & morphology & $R_\text{proj}$ & log($M_\ast$/$M_\sun$) & $N_{\text{S/N} \geq 8,\,r > 15\arcsec}$  & $<R_\text{GCs}>$\\ 
	& 	&  & (Mpc) & & & ($R_\text{eff}$) \\
(1) & (2) & (3) & (4) & (5) & (6) & (7)\\
\hline\hline
FCC083 & NGC\,1351 &  E5 & 0.58 & 10.5 & 15 &  0.70 \\
FCC090 & PGC\,13058 & E4 & 0.57 & 8.9 & 1 & 1.11  \\
FCC113 & ESO\,358-015 & Scd & 0.43 & 8.3 & 1 & 0.75 \\
FCC143 & NGC\,1373 &  E3 & 0.26 & 9.4 & 3 & 1.91 \\
FCC147 & NGC\,1374 &E0 & 0.22 & 10.4 & 18 & 1.20 \\
FCC148 & NGC\,1375 & S0 & 0.22 & 9.8 & 2 & 2.10 \\
FCC153 & IC\,1963 & S0 &  0.40 & 9.9 & 1 & 1.26 \\
FCC161 & NGC\,1379 & E0 & 0.17 & 10.4 & 27 & 0.97 \\
FCC167 & NGC\,1380 & S0 & 0.21 & 11.0 & 16 & 0.68 \\
FCC170 & NGC\,1381 &  S0 & 0.14 & 10.4 & 9 & 2.00 \\
FCC176 & NGC\,1369 & SB & 0.30 & 9.8 & 1 & 0.41 \\
FCC177 & NGC\,1380A & S0 & 0.27 & 9.9 & 6 & 0.68 \\
FCC182 & -- & SB0 & 0.11 & 9.2 & 2 & 1.74 \\
FCC184 &  NGC\,1387 &  SB0 & 0.11  & 10.7 & 18 & 1.25 \\
FCC190 & NGC\,1380B  & SB0 & 0.13 & 9.7 & 9 & 1.22 \\
FCC193 & NGC\,1389 & SB0 & 0.13 &  10.5 & 1 & 0.72 \\
FCC213 & NGC\,1399 & E1 &  0 & 11.4 & 25 & 0.12 \\
FCC219 & NGC\,1404 & E2 & 0.06 & 11.1 & 3 & 0.22 \\
FCC249 & NGC\,1419 & E0 & 0.71 & 9.7 & 3 & 1.72\\
FCC255 & ESO\,358-G50 & S0 & 0.60 & 9.7 & 3 & 1.38 \\
FCC276 & NGC\,1427 & E4 & 0.27 & 10.3 & 19 & 0.77 \\
FCC290 &  NGC\,1436 & Sc & 0.38 & 9.8 & 1 & 0.34  \\
FCC308 & NGC\,1437B & Sd & 0.60 & 8.6 & 3 & 1.40 \\ \hline
\end{tabular}
\tablefoot{(1) Galaxy name from \cite{Ferguson1989} and (2) alternative name. (3) galaxy morphology. (4) Projected distance from FCC\,213. (5) Stellar mass from \cite{Iodice2019} and \cite{Liu2019}. (6) Number of GCs with S/N $>$ 8 and galactocentric distance $>$ 15\arcsec. (7) Mean galactocentric distance of the F3D GCs. This refers to all GCs (see paper I).}
\end{table*}

\section{Extraction of stellar population properties}
\label{sect:stellar_pop_methods}
We describe how the stellar population properties were derived from the GC spectra in the following section. Besides the default approach of measuring metallicities discussed in the main text, we tested other approaches to derive metallicities as described in App. \ref{app:SSP_models}.

\subsection{Full spectral fitting with pPXF}
\label{sect:fitting_ppxf}
We used the penalised Pixel-fitting (\textsc{pPXF}) method \citep{Cappellari2004, Cappellari2017} to obtain metallicities of GCs with S/N $\geq$ 8 \AA$^{-1}$. \textsc{pPXF} uses a penalised maximum likelihood approach to fit spectra with a combination of user-provided template spectra and allows to determine best-fit age and metallicity distributions from a library of single stellar population (SSP) models (e.g. \citealt{Pinna2019, Boecker2019, Fahrion2019, Fahrion2019b}). We used the E-MILES SSP models \citep{Vazdekis2016}, that have broad wavelength coverage from 1680 to 50000\,\AA\,. The model spectra are sampled at 1.25\,\AA\,at a spectral resolution of $\sim$\,2.5\,\AA\,\citep{FalconBarroso2011} in the wavelength range of interest, approximately corresponding to the mean instrumental resolution of MUSE. We used additive polynomials of degree 12 for the extraction of LOS velocities and multiplicative polynomials of degree 8 for the stellar population measurements.

Throughout this work, we used SSP models based on BaSTI isochrones \citep{Pietrinferni2004, Pietrinferni2006} and a MW-like double power law (bimodal) initial mass function with a high mass slope of 1.30 \citep{Vazdekis1996}. The models provide a grid of SSPs with ages between 300 Myr and 14 Gyr and metallicities between \mbox{[M/H] = $-2.27$ and $+0.40$ dex}. 
Because \textsc{pPXF} returns the weights of the best-fitting combination of SSP models, the stellar populations of a GC can be described by the weighted mean age and metallicity. 
As described in paper I, we fitted each GC spectrum in Monte-Carlo-approach to derive reliable random uncertainties by perturbing the spectrum 100 times based on the residuals from the first fit. These fits were done with a restricted library that only contains SSP templates with stellar ages $\geq$ 8 Gyr. This limits the effect of a possible age-metallicity degeneracy and speeds up the fitting process. To explore the influence of this choice, we also fitted a sub-sample of 135 GCs with S/N $>$ 10\,\AA\, without any constraint on the age. The results of the GC ages are described in Sect. \ref{sect:ages}. 

To determine reliable metallicities from the GC spectra, we used the E-MILES SSP models because their broad wavelength coverage helps to reduce uncertainties. 
For E-MILES, only so-called baseFe models are available that are based on empirical stellar spectra and thus inherit the abundance pattern of the MW. They are $\alpha$-enhanced at low metallicities and follow [Fe/H] = [M/H] at high metallicities. As this abundance pattern might not represent the GCs in the Fornax cluster, we determined the [$\alpha$/Fe] abundances of a sub-sample of GCs with the highest S/N ($>$ 20 \AA$^{-1}$) using $\alpha$-variable MILES models that are based on the standard MILES models which only offer two different [$\alpha$/Fe] values of 0 (scaled solar) and 0.4 dex ($\alpha$-enhanced). The $\alpha$-variable MILES models were created using a linear interpolation between these to create a regular grid from [$\alpha$/Fe]  = 0 to [$\alpha$/Fe]  = 0.4 dex with a spacing of 0.1 dex. 
Because these $\alpha$-variable MILES models introduce another free parameter that can be fitted, only GCs with the highest S/Ns give reasonable results. The measured abundance pattern (see Sect. \ref{sect:iron_and_total_metallicities}) of the high S/N GCs further supports the use of the E-MILES models as our default approach.


To summarise, our default approach was to fit the GCs with S/N $\geq$ 8 \AA$^{-1}$ with the E-MILES SSP models and an age constraint of $\geq$ 8 Gyr. We tested the effects of GC age by fitting GCs with S/N $>$ 10 \AA$^{-1}$ without age constraint and determined [$\alpha$/Fe] abundances only for the brightest GCs with S/N $>$ 20 \AA$^{-1}$. The latter two approaches are to validate the results from our default approach. In App. \ref{app:SSP_models}, we further explore the choice of SSP models and also test metallicities from line-strength indices.

\subsection{Globular cluster colours}
\label{sect:colours}
We used ($g - z$) colours, mostly from the photometric GC catalogues of \cite{Jordan2015} that were obtained as part of the ACS Fornax Cluster Survey (ACSFCS; \citealt{Jordan2007}). These catalogues report the magnitudes of the GC candidates in the ACS F475W ($\sim g$ band) and F850LP ($\sim z$ band). 

Not all galaxies in our sample were covered by the ACSFCS and consequently, 45 GCs in our sample have no ACS colours available. For those, we determined synthetic ($g - z$) colours from the MUSE spectra using the F475W and F850LP transmission curves. While the F475W band is covered completely with MUSE, the F850LP bandpass extends outside the MUSE coverage. In the colour regime covered by the 45 GCs without ACS photometry, the synthetic colours agree with the ACSFCS colours within a scatter of $\sim$ 0.05 mag. 

\section{Results}
\label{sect:results}
In the following, we present the results from the stellar population analysis of the F3D GCs. We first discuss the colour-metallicity relation (CZR) and then address GC ages, $\alpha$-abundances and the relation between mass and metallicity.

\subsection{Colour metallicity relation}
\label{sect:CZR}

\begin{figure}
\centering
\includegraphics[width=0.49\textwidth]{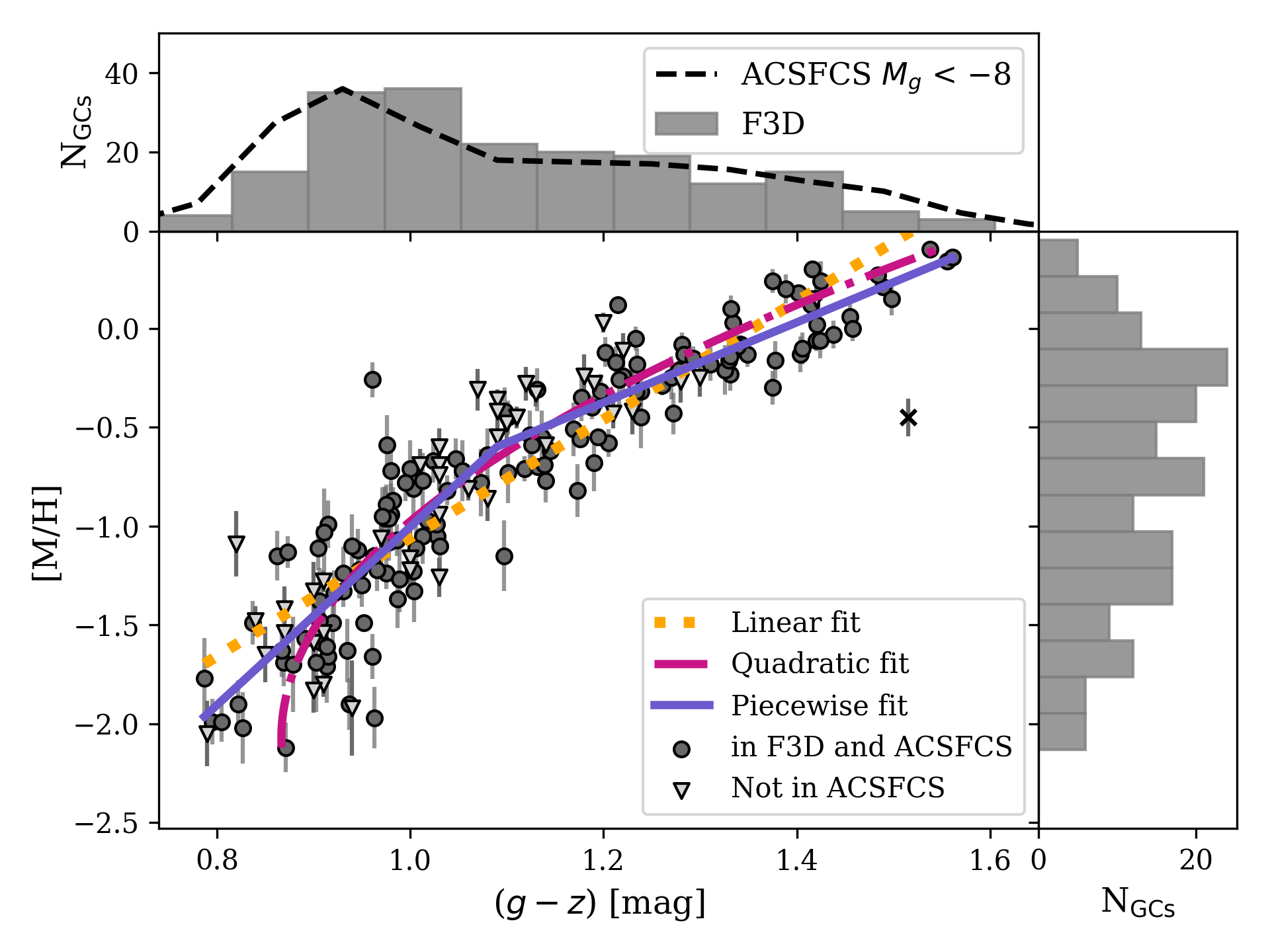}
\caption{Colour-metallicity distribution of F3D GCs. Spectroscopic GC metallicities from full spectral fitting versus ($g - z$) colour. Filled circles and triangles refer to GCs with and without ACSFCS ($g - z$) colours from \cite{Jordan2015}, respectively. For the latter, we obtained the colour from the MUSE spectrum directly. The orange, pink and purple lines give the fit using the linear, quadratic or piecewise function (Eq. \ref{eq:lin_CZR}, \ref{eq:quad}, and \ref{eq:piecewise}). The corresponding residuals are found in Fig. \ref{fig:residuals}. As described in the text, we excluded an outlier from the fit marked by a cross. Separate distributions of colour and metallicities are shown on the top and the right-hand side. In the top panel, we included the histogram from the full ACSFCS GC sample after applying a magnitude cut corresponding to our sample.}
\label{fig:metallicity_vs_color}
\end{figure}

\begin{figure}
\centering
\includegraphics[width=0.49\textwidth]{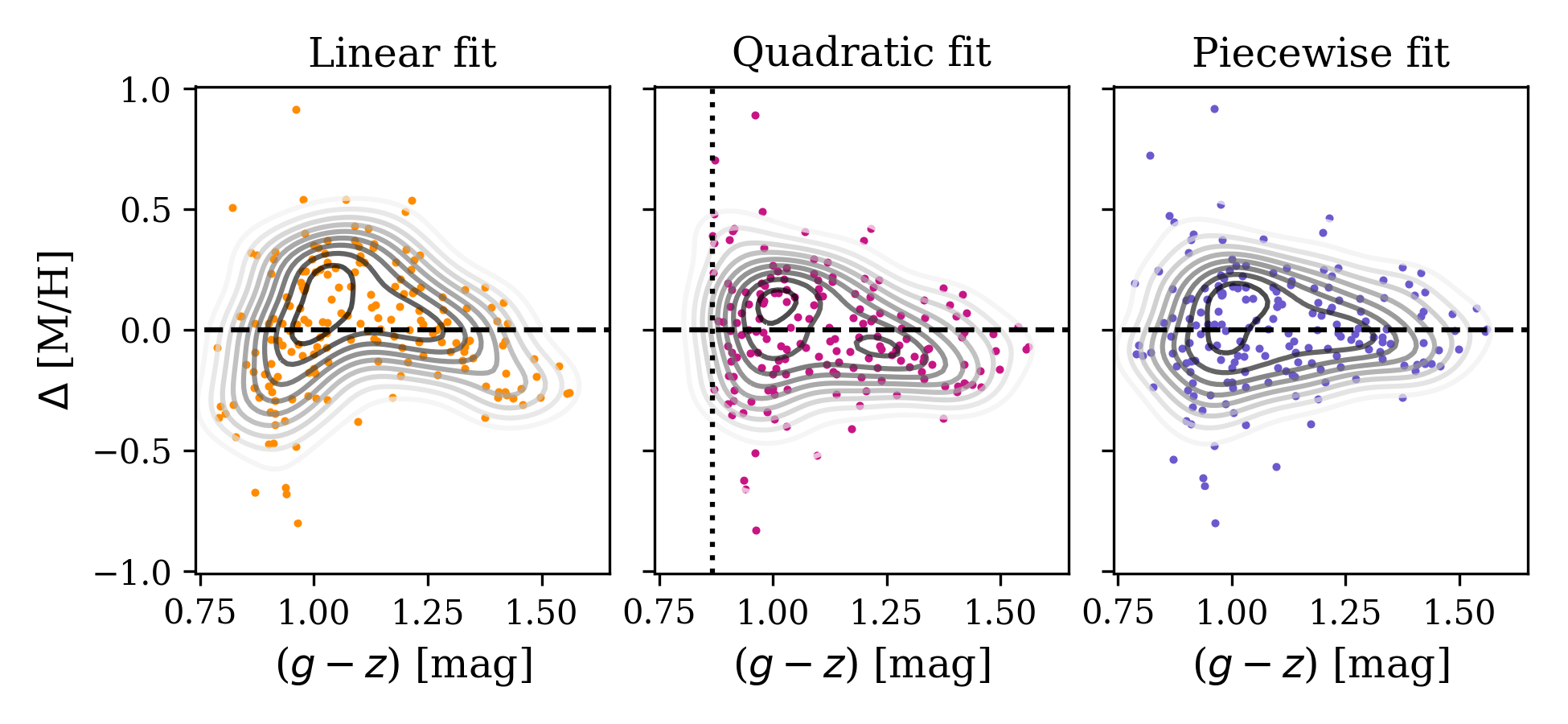}
\caption{Residuals when fitting the colour-metallicity distribution shown in Fig. \ref{fig:metallicity_vs_color} with a linear (\textit{left}), a quadratic (\textit{middle}) and piecewise linear function (\textit{right}), respectively. Coloured points show the observed scatter and the contours give a kernel density estimation using an arbitrary Gaussian kernel for visualisation of the residual shape. The dotted vertical line in the middle panel shows where the quadratic model diverges.}
\label{fig:residuals}
\end{figure}

We plot the distribution of colours and metallicities of our F3D GCs in Fig. \ref{fig:metallicity_vs_color}. This figure shows the relation between the ($g - z$) colours and spectroscopic total metallicity [M/H] as derived based on the E-MILES SSP models for the GCs with S/N $\geq$ 8 \AA$^{-1}$. Although most of the GCs were covered by the ACSFCS and have HST colours, the CZR is better constrained when also including GCs with synthetic MUSE colours. There is one outlier with a ACSFCS colour of ($g - z$) $\sim$ 1.5 mag and a metallicity of \mbox{$\sim -0.5$} dex that lies significantly below the relation. This is a GC found in the halo pointing of FCC\,167 with a synthetic MUSE colour of ($g - z$) $\sim$ 1.2 mag that would place it among the bulk of GCs. Usually, the synthetic colours agree within $\pm 0.05$ mag with the ACSFCS colours, making this GC an outlier and because the origin of the large colour difference is unknown, we excluded this GC from the fit. Another visible outlier that lies above the relation at ($g - z$) $\sim$ 1.0 is a GC found in the central pointing in FCC\,276 with a small galactocentric distance of 16\arcsec. Since FCC\,276 is quite massive (log($M_\ast/M_\sun) \sim 10.3$), it is possible that the spectrum of this GC is still contaminated by the bright galaxy background which could bias the measured metallicity to higher values.

The top panel of Fig. \ref{fig:metallicity_vs_color} compares the colour histogram of the F3D GCs to the full sample of ACSFCS GCs \citep{Jordan2015}, normalised to match the peak in our GC distribution. 
We only have metallicity estimates from GCs with spectral \mbox{S/N $\geq$ 8 \AA$^{-1}$}. As we showed in paper I, these are GCs with \mbox{$M_g \lesssim -8$ mag}. Therefore, we apply the same brightness cut to the full ACSFCS sample. Our GC sample is representative of the bright GC population of the ACSFCS cluster survey, and the full colour range from 0.8 to 1.6 mag is well sampled. There is a large number of GCs with $(g - z) \sim 1$ mag, but our sample shows a deficit of GCs at very blue colours $<$ 0.8 mag, possibly because those are expected to be very metal-poor and consequently the absence of strong absorption lines in the spectrum leads to lower S/N. 

Our GC sample contains the most massive GCs of the total population and in order to apply our relation to the full GC distribution (see Sect. \ref{sect:MDFs}), we have to assume that the less massive GCs follow the same relation. In Sect. \ref{sect:MZR}, we report on the mass-metallicity relation (MZR) of GCs and show that the metallicity does not depend strongly on the GC stellar mass. The less massive GCs missing from our sample are expected to be even more metal-poor and thus it is unlikely that they would change the shape of the CZR. In addition, the colour span around ($g - z$) $\sim 1$ mag, where we observe the break in the relation, is already well sampled.

To quantitatively describe the CZR, we fitted the distribution with different functions using a least-square algorithm. The best-fitting functions are shown as coloured lines in Fig. \ref{fig:metallicity_vs_color} and we show the respective residuals in Fig. \ref{fig:residuals}.
Using a simple linear function gives a relation of the following form:
\begin{equation}
[\text{M/H}] =  (-4.05 \pm 0.11) + (2.99 \pm 0.10)\,(g - z),
\label{eq:lin_CZR}
\end{equation}
As Fig. \ref{fig:residuals} shows, the residual of this linear fit shows a bent shape. At very blue and red colours, the metallicities are overestimated and are underestimated at intermediate colours. 

In order to improve the quantitative description of the CZR, we used a quadratic relation to fit the CZR (see also \citealt{Sinnott2010, Harris2017}):
\begin{equation}
(g - z) = a\,\text{[M/H]}^2 + b\,\text{[M/H]} + c.
\label{eq:quad}
\end{equation}
The least-square fit returned best-fitting paramters of \mbox{$a = 1.34 \pm 0.01$}, \mbox{$b=0.46 \pm 0.02$}, \mbox{$c=0.11 \pm 0.01$}. This best-fitting relation is shown by the pink line in Fig. \ref{fig:metallicity_vs_color}. The residual shows a more symmetric shape than when using the linear fit.

In addition, we used a piecewise linear function, similar to that of \cite{Peng2006}:
\begin{equation}
\label{eq:piecewise}
\begin{split}
\text{[M/H]} & = b_1 + m_1 (g - z)\,\hspace{0.5cm}\text{for}\,(g - z) < x_0 \\
 	& = b_2 + m_2 (g - z)\,\hspace{0.5cm}\text{for}\,(g - z) \geq x_0,
\end{split}
\end{equation}
with best-fitting parameters of \mbox{$m_1 = 4.51 \pm 0.32$}, \mbox{$b_1 = -5.51 \pm 0.36$}, \mbox{$m_2 = 2.03 \pm 0.20$}, \mbox{$b_2 = -2.81 \pm 0.36$}, and \mbox{$x_0 = 1.09 \pm 0.03$} (purple line in Fig. \ref{fig:metallicity_vs_color}). The residual is more symmetric around the zero line (Fig. \ref{fig:residuals}). 

Comparing the residuals of the fitted relations shows that the linear fit is insufficient to capture the shape of the CZR accurately. The quadratic and piecewise relations return similar residuals, however, the quadratic relation shows an asymptotic behaviour for colours ($g - z$) $<$ 0.86 mag, although our sample reaches bluer colours. To compare the models quantitatively, we derived the Bayesian information criterion (BIC) for each model. The piecewise linear relation has the lowest BIC and is preferred over the linear model by $\Delta$BIC = 32 and over the quadratic model by $\Delta$BIC = 35. While the residual scatter is comparable for the piecewise and the quadratic model, the asymptotic behaviour of the latter reduces the number of observables and thus increases the BIC. We conclude that the piecewise relation best represents the data.

\subsection{Globular cluster ages}
\label{sect:ages}
\begin{figure}
\centering
\includegraphics[width=0.49\textwidth]{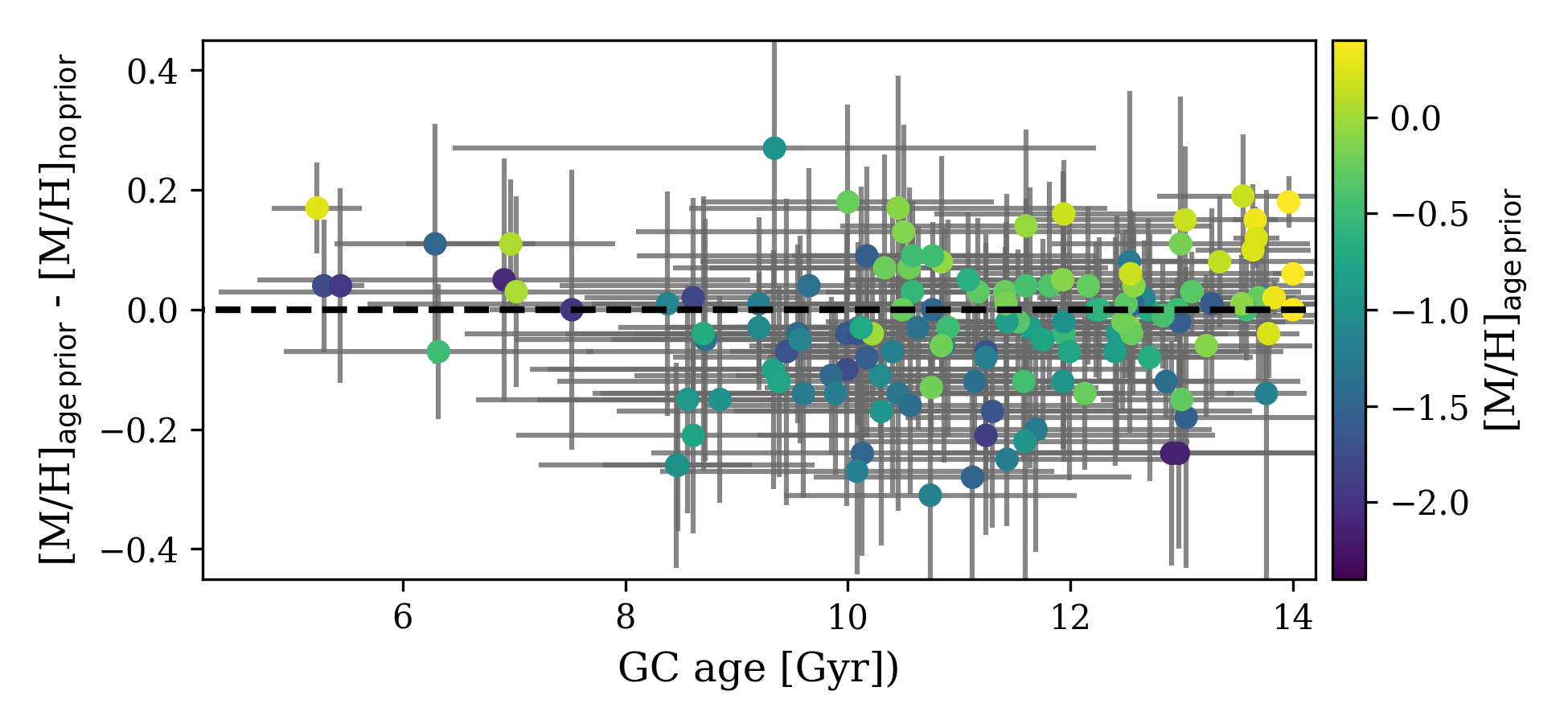}
\caption{Difference of GC metallicities from \textsc{pPXF} when using or not an age prior ($\geq$ 8 Gyr) in relation to the best-fitting age. The symbols are colour-coded by the best-fitting metallicity. The dashed line shows the zero difference.}
\label{fig:age_comp}
\end{figure}

\begin{figure}
\centering
\includegraphics[width=0.49\textwidth]{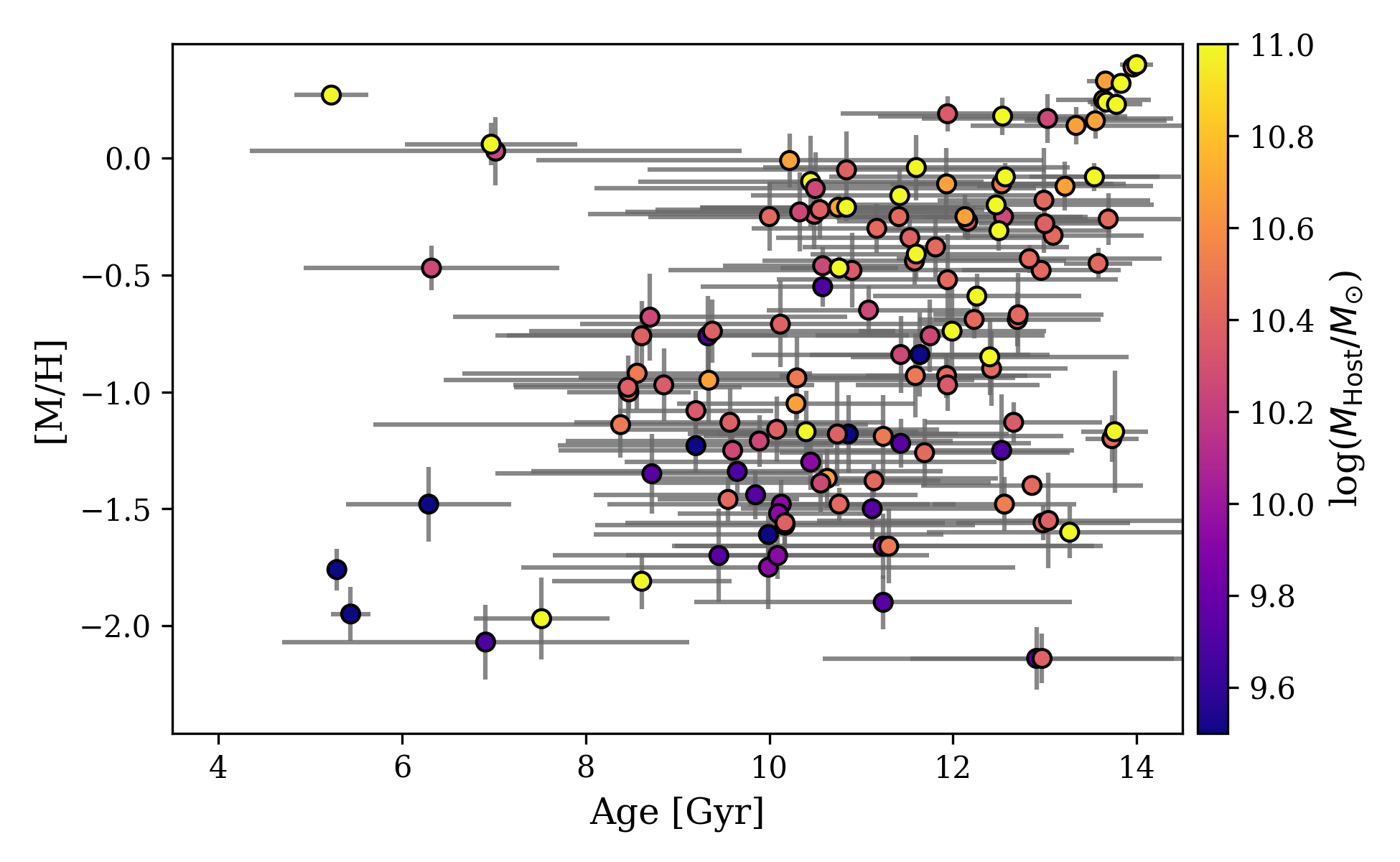}
\includegraphics[width=0.49\textwidth]{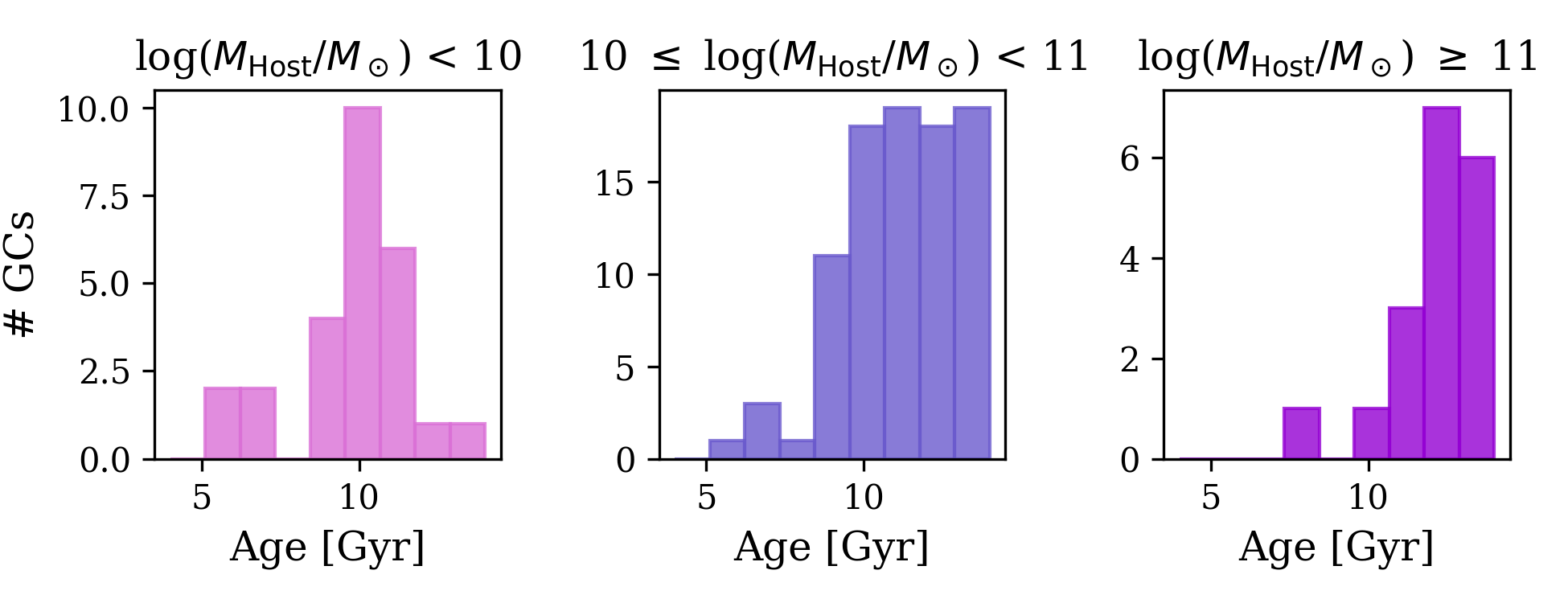}

\caption{\textit{Top:} Best-fitting ages and metallicities of GCs, inferred from full spectral fitting with the E-MILES templates and no age constraint. The colour coding refers to the stellar mass of the host galaxy \citep{Iodice2019, Liu2019}. \textit{Bottom:} GC age distributions for host galaxies in three different mass bins.}
\label{fig:ages}
\end{figure}

While our standard approach for fitting the GC spectra assumes an age $\geq$ 8 Gyr, we also fitted a sub-sample of 135 GCs with high S/N without any constraints on the age. Because of the larger SSP model grid, these fits take substantially longer, but allow us to study the effect of GC ages on the CZR due to a possible age-metallicity degeneracy.
Therefore, in Fig. \ref{fig:age_comp}, we plot the GC metallicities from the default approach (with age prior) and without age constraint as a function of the best-fitting age. The metallicities from both methods agree within the uncertainties and there seems to be no trend with age. With very few exceptions, the GCs show best-fitting ages $\geq$ 8 Gyr, validating our choice of restricting the model grid for the \textsc{pPXF} fit. 

Consequently, fitting without age constraint results in a similar non-linear CZR as is shown in Fig. \ref{fig:colour_met_relation_comp} in the Appendix with best-fitting parameters presented in Tab. \ref{tab:colour_met_results}. We can therefore conclude that the shape of the CZR cannot be explained by an underlying age-metallicity degeneracy.

The reddest, most metal-rich GCs in the sampe have very small age and metallicity uncertainties. For them, it is likely that the small uncertainties are an effect of the limited SSP grid. Otherwise, the GC ages have typical random uncertainties of $>$ 2 Gyr, reflecting the challenging age determination of old stellar populations (e.g. \citealt{Usher2019}, or App. in \citealt{Fahrion2019b}). The wavelength coverage of MUSE is further lacking age sensitive spectral features such as higher Balmer lines. 

The upper panel in Fig. \ref{fig:ages} shows the age-metallicity distribution of the F3D GCs, colour-coded by the stellar mass of the host \citep{Liu2019, Iodice2019}. This figure suggests a shallow age-metallicity correlation of the GCs in which more metal-rich GCs are also older. This trend is mainly driven by the reddest, most metal-rich GCs that show very small age and metallicity uncertainties. As mentioned, it is likely that these GCs exceed the metallicities of the SSP models, or are strong $\alpha$-enhanced, as was found for several GCs of massive ETGs (e.g. \citealt{Puzia2005, Puzia2006, Woodley2010b}). The other GCs show a very mild correlation between age and metallicity that also coincides with a relation between host mass and GC age. 

To illustrate this, we binned the GC sample based on the stellar mass of the host galaxy into three mass bins and the bottom panel of Fig. \ref{fig:ages} shows the GC age distribution in these mass bins. The GCs with the lowest host masses (log($M_\text{Host}/M_\sun$) $<$ 10) show a peak at 10 Gyr, whereas the intermediate and high mass bins show distributions that are dominated by very old ages. While the intermediate mass bin (10 $<$ log($M_\text{Host}/M_\sun$) $<$ 11) shows some GCs with ages $<$ 10 Gyr, these slightly younger GCs are apparently missing in the highest mass bin (log($M_\text{Host}/M_\sun$) $>$ 11). 
Although the the number of GCs in each mass bin is quite low and the age uncertainties are large, we found indications that the lower mass hosts indeed have younger GC systems. Indications for such a trend were also found, for example, by \cite{Usher2019} when comparing three SLUGGS galaxies, possibly due to a top-down formation of GCs that form later in less massive galaxies. Moreover, such a behaviour is in agreement with a mass-dependent age-metallicity relation (see \citealt{Leaman2013a, Boecker2020} and references therein). Because low mass galaxies also tend to have more metal-poor GCs, the observed weak age-metallicity correlation might be driven by the host mass.

\begin{figure*}
\centering
\includegraphics[width=0.99\textwidth]{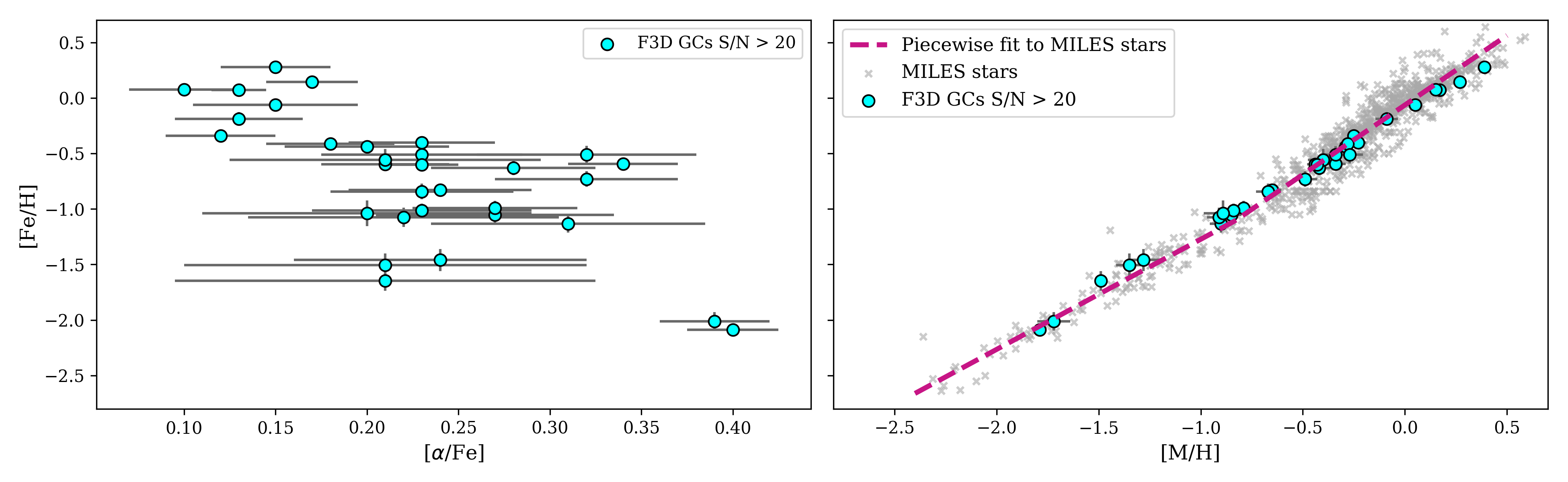}
\caption{Abundance pattern of F3D GCs. [Fe/H] in relation to [$\alpha$/Fe] (\textit{left}) and [M/H] (\textit{right}). In the right panel, the grey crosses show the distribution for the MILES stars that were used to construct the E-MILES models and the purple line is a least-square fit with a piecewise linear function (Eq. \ref{eq:piecewise}) to the MILES stars. The cyan circles show the brightest F3D GCs in our sample.}
\label{fig:alpha_variable}
\end{figure*}

\subsection{Abundance pattern of GCs}
\label{sect:iron_and_total_metallicities}
We obtained [$\alpha$/Fe] values for the 31 brightest GCs in our sample that have S/N $>$ 20 \AA$^{-1}$ using full spectral fitting with $\alpha$-variable MILES models. We show the [$\alpha$/Fe] values in relation to the iron metallicities of these GCs in the left panel of Fig. \ref{fig:alpha_variable}. This figure illustrates that these GCs show an negative correlation between metallicity and $\alpha$-abundance with the metal-poor GCs being more $\alpha$-enhanced. 

The GCs therefore seem to follow a similar abundance pattern as the MW stars used to create the E-MILES SSPs. This is shown in the right panel of Fig. \ref{fig:alpha_variable}, where we plot the relationship between iron and total metallicity for the MILES stars and the F3D GCs. Although the sample of GCs is limited, they seem to follow the same trend. This indicates that the E-MILES SSP models are indeed a reasonable choice to use with the GC spectra.

For this reason, we used the [M/H]-[Fe/H] relation of the MILES stars to establish a conversion between total and iron metallicities for the E-MILES GC metallicities.
The purple line in Fig. \ref{fig:alpha_variable} shows a least-square piecewise fit to the MILES stars (Eq. \ref{eq:piecewise}) with parameters: $m_1$ = 0.99 $\pm$ 0.03, $b_1$ = $-$0.28 $\pm$ 0.17, $m_2$ =  1.25 $\pm$ 0.02, $b_2$ = $-$0.06 $\pm$ 0.18, and \mbox{$x_0$ = $-$0.83 $\pm$ 0.11}.

\subsection{Mass-metallicity relation}
\label{sect:MZR}
\begin{figure}
\includegraphics[width=0.49\textwidth]{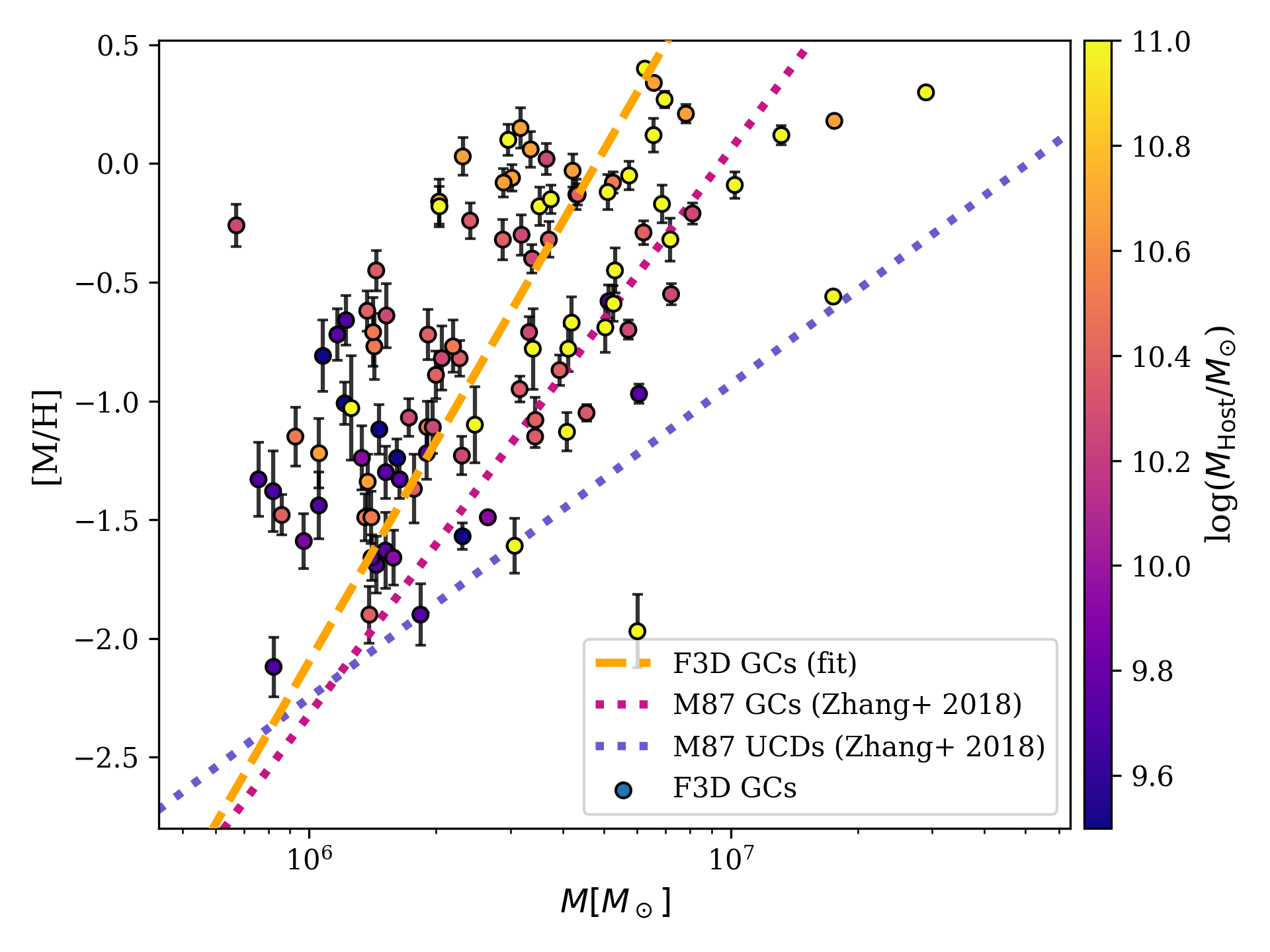}
\caption{MZR for the F3D GCs in comparison to the MZR for GCs and UCDs in M87 (pink and purple dotted lines, respectively, \citealt{Zhang2018}). The least-square fit to our data is shown in orange. The F3D GCs are colour-coded by the stellar mass of their host.}
\label{fig:mass_metal_relation}
\end{figure}

We determined the stellar masses of all GCs with ACSFCS colours using their measured metallicities and the photometric predictions from the E-MILES models that give the stellar mass-to-light ratio for a given model\footnote{\url{http://research.iac.es/proyecto/miles//pages/photometric-predictions-based-on-e-miles-seds.php}}. Assuming an average distance to the Fornax cluster of 20.9 Mpc \citep{Blakeslee2009}, we converted the $g$-band magnitudes from the ACSFCS \citep{Jordan2015} to luminosities and then translated those to stellar masses. This results in the MZR shown in Fig. \ref{fig:mass_metal_relation}. We found GC masses between a few 10$^{5} M_\sun$ and a few 10$^{7} M_\sun$, representative of the more massive GC population. In this figure, the MZR of the GCs is compared to a MZR for GCs of M87 \citep{Zhang2018}. They also reported the MZR for ultra compact dwarfs (UCDs) which is shallower than that of the GCs. 

We fitted a log-linear function to describe the MZR:
\begin{equation}
\text{[M/H]} = -20.58 + 3.08 \, \text{log}(M_{\ast, \text{GC}}/M_\sun)
\label{eq:MZR}
\end{equation}

We found indications that the more massive galaxies have more massive GCs at the same metallicity. This could also explain the offset with respect to the relation from \citep{Zhang2018} because M87 is significantly more massive than the galaxies included in our sample. However, the steepness of this relation shows that the metallicity of the GCs is not influenced strongly by the stellar mass of the GC, although there is a weak correlation. The less massive GCs tend to be more metal-poor and consequently, it is unlikely that including fainter GCs into the CZR would change the shape. 

The MZR can also give insights into the origin of the so-called blue tilt, an observed optical colour-magnitude relation of blue GCs that describes that the brighter GCs of the blue GC population tend to be redder (e.g \citealt{Harris2006, Spitler2006, Strader2006, Mieske2006}). This has often been interpreted as an result of an underlying MZR of GCs, where the more massive GCs can retain more metals and thus have redder colours (e.g. \citealt{StraderSmith2008, BailinHarris2009}). Recently, \cite{Usher2018} used simulations to explore the origin of the blue tilt and suggested that its origin lies in the lack of massive metal-poor GCs because those would require special formation conditions with high gas densities in a metal-poor environment. The weak MZR we found is in accordance with this picture.  

\section{Discussion}
\label{sect:discussion}
We discuss our derived CZR with the literature as follows. We also explore the dependence of the CZR on the stellar mass of the host and discuss the implications of our findings.

\subsection{Comparison to literature}
\label{sect:literature}
\begin{figure}
\centering
\includegraphics[width=0.49\textwidth]{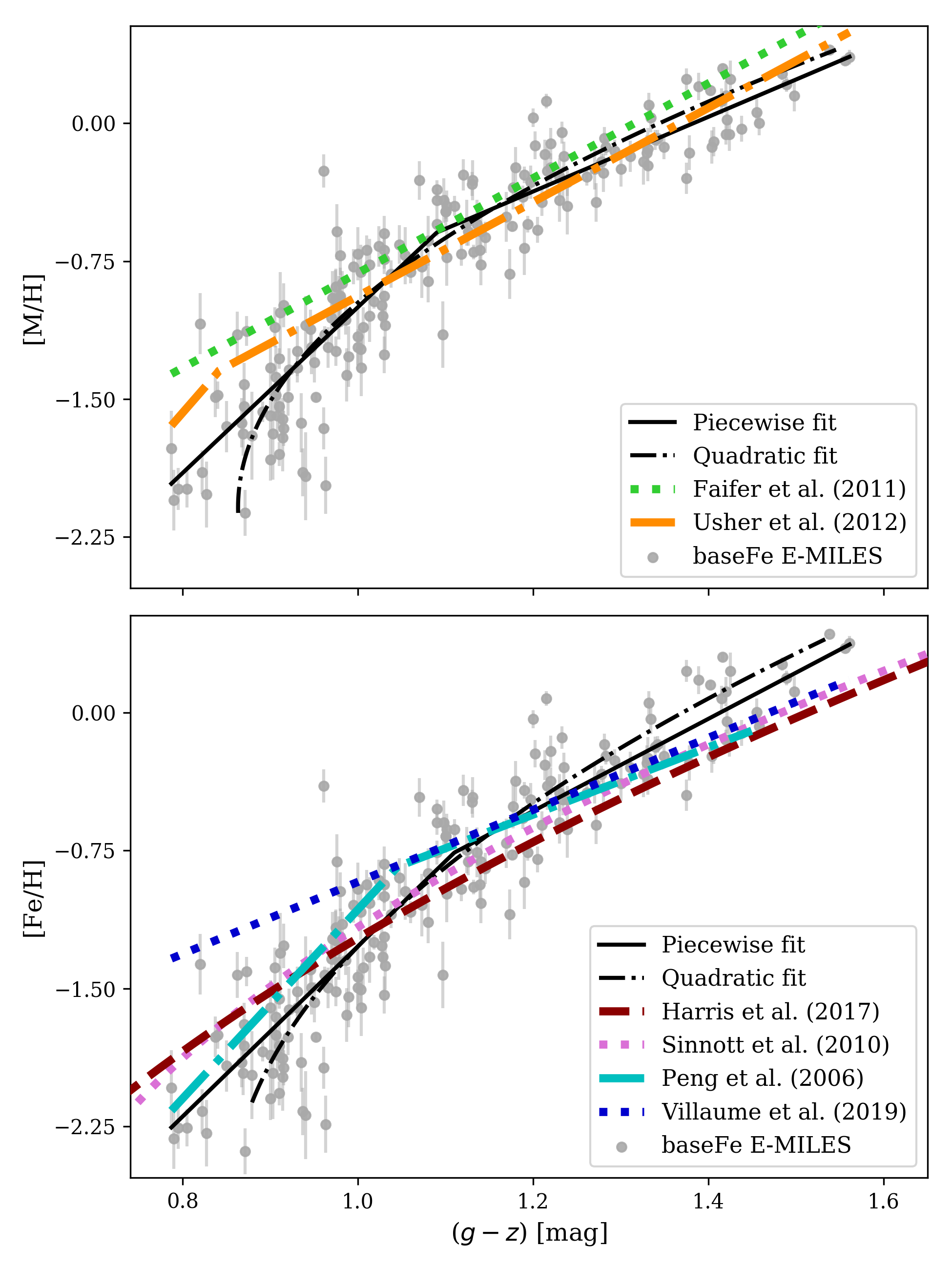}
\caption{Our CZR in comparison to literature results. We differentiate between relations based on total metallicities (\textit{top}) and iron metallicities (\textit{bottom}). The black curves give our fits (Tab. \ref{tab:colour_met_results}). The green line in the top plot refers to the relation of \cite{Faifer2011} of five massive ETGs and the orange line shows the piecewise relation of \cite{Usher2012}. Both relations were converted from ($g - i$) colours to ($g - z$) using the translations given in \cite{Usher2012}. In the bottom panel, we show the quadratic relations of \cite{Sinnott2010} and \cite{Harris2017} (pink and red lines) obtained for GCs of Centaurus\,A and the linear relation of \cite{Villaume2019} based on metallicities of GCs of M87 (dark blue line). The bright blue line shows the piecewise relation from \cite{Peng2006} obtained from a diverse sample of literature metallicities.}
\label{fig:CZR_literature}
\end{figure}

Our CZR is compared to relations from the literature in Fig. \ref{fig:CZR_literature}, shown by lines of different colours. We differentiate between relations based on total and iron metallicities to avoid further conversions between them. For our sample, we used the conversion derived in Sect. \ref{sect:iron_and_total_metallicities} to convert them from [M/H] to [Fe/H]. The best-fit parameters of the CZR based on iron metallicities can be found in Table \ref{tab:colour_met_results}.

\cite{Peng2006} studied the bimodality of GC colours in the Virgo cluster using HST/ACS photometry and ($g - z$) colours. They derived a CZR from the few spectroscopic GC metallicity measurements of the MW, M87 and M49 that were available at that time \citep{Harris1996, Cohen1998, Cohen2003}. Their CZR is described by a piecewise linear relation with a break at ($g - z$) $\approx$ 1.05 mag. As Fig. \ref{fig:CZR_literature} shows, their relation is close to ours, especially for the blue GCs. At redder colours their relation is shallower. As \cite{Villaume2019} discussed, the break in the \cite{Peng2006} relation might be mainly caused by the MW GCs that are significantly more metal-poor than those of M87. 

\cite{Faifer2011} studied the GC systems of five massive ETGs with photometry in the $g$ and $i$-bands of the Gemini Multi-Object Spectrograph. Using literature metallicities \citep{Pierce2006a, Pierce2006b, Norris2008}, they found a linear CZR which is shown by the green line in Fig. \ref{fig:CZR_literature}. We converted their ($g^\prime - i^\prime$) colours to ($g - z$) using the translation given in the Appendix of \cite{Usher2012}. Their relation predicts higher GC metallicities at all colours, although the slope is very similar to that of \cite{Usher2012}, who used literature metallicities from \cite{Kuntschner2002, Brodie2005, Cenarro2007, Chomiuk2008, Caldwell2011} and SLUGGS ($g - i$) photometry, to derive a piecewise CZR shown by the orange line in Fig. \ref{fig:CZR_literature}. This relation fits the red GCs of our sample quite well, but the break point is located at colour of \mbox{($g - z$) $\approx$ 0.84 mag}. The position of the break point is strongly driven by the metallicities of M31 GCs \citep{Caldwell2011} because the other galaxies in this collection show no GC metallicities $< -1.2$ dex. In the sample, M31 is also the only LTG, while the others are massive ETGs. 

Using metallicities from \cite{Woodley2010} and $griz$ photometry, \cite{Sinnott2010} presented a quadratic CZR for GCs of the giant elliptical Centaurus A. \cite{Harris2017} used the same metallicities, but combined the $griz$ photometry of \cite{Sinnott2010} with $UBVRI$ photometry available from \cite{Peng2004} to derive a very similar quadratic relation using ($g - I$) colours. They also give conversion to ($g - z$) colours (see also \citealt{Choksi2019}). Their CZR is offset to our red GCs and shallower at blue colours.

Very recently, \cite{Villaume2019} presented a sample of 177 GCs of M87 with spectroscopic metallicities and found a linear relation shown in Fig. \ref{fig:CZR_literature}. Their CZR follows the relation of \cite{Harris2017} at red colours and shows a deviation from our relation at the bluest GC colours. \cite{Villaume2019} discussed that their findings of more metal-rich blue GCs could indicate an environmental effect caused by the assembly history of M87 itself.

The comparison to literature CZR highlights the diversity of relations that were found using different techniques and studies of different environments. In general, it appears that studies focusing on massive ETGs generally find linear relations due to a lack of metal-poor GCs. Non-linear relations are predominantly found when incorporating measurements of metal-poor GCs, for example, from the MW or M31. This could indicate that massive ETGs indeed have a different CZR, whereas it is also possible that the lack of metal-poor GCs is due to selection effects and limited sample sizes because the most massive galaxies are dominated by more metal-rich GCs. Additionally, the radial extent of the studied GCs can bias the selection as the blue GC population usually is more extended (e.g. \citealt{Harris2016}) and thus concentrating on the inner regions of massive galaxies can result in a lack of blue GCs. However, because our sample uses a large variety of galaxy masses, also the metal-poor end of the CZR is well sampled.

\begin{figure}
\centering
\includegraphics[width=0.49\textwidth]{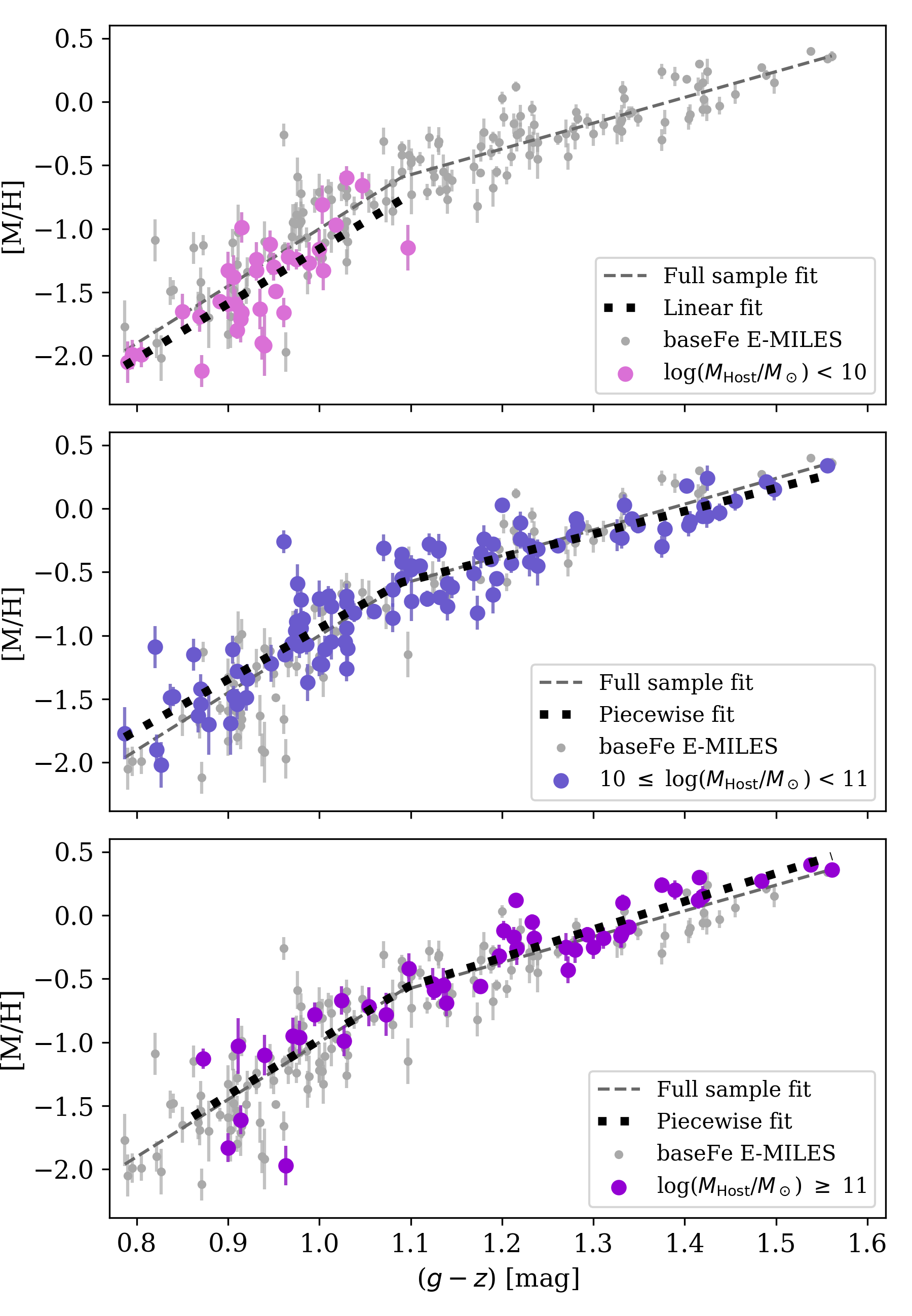}
\caption{CZR for GCs that are hosted by galaxies of different stellar masses. We binned the sample into three mass bins. \textit{Top:} Pink circles show GCs for galaxies with stellar masses log($M_\text{Host}/M_\sun$) $<$ 10. These galaxies are the least massive ones in our sample and generally have blue, metal-poor GCs. \textit{Middle}: Coloured circles show GCs for host masses with 10 $\leq$ log($M_\text{Host}/M_\sun$) $<$ 11. This bin contains the most GCs and those sample the full range of colours and metallicities. \textit{Bottom}: Highest mass bin with log($M_\text{Host}/M_\sun$) $\geq$ 11 containing GCs with a broad range of metallicities and colours, except for the most metal-poor ones. The grey dots and line give the full sample and the fit, respectively. The black dotted line shows the fit to the respective mass bin.}
\label{fig:mass_comp}
\end{figure}

\begin{figure*}
\centering
\includegraphics[width=0.99\textwidth]{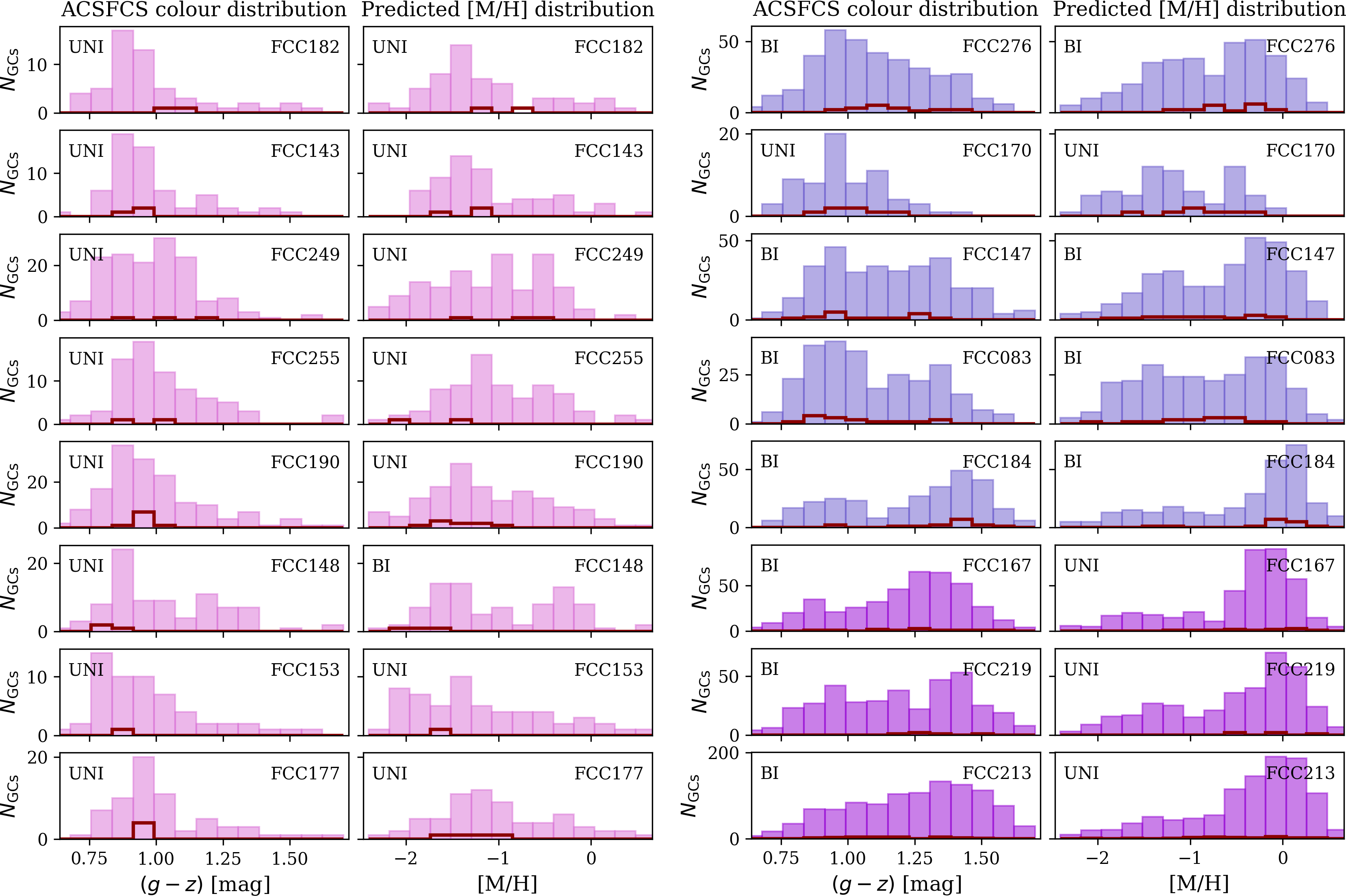}
\caption{Colour (first and third panel) and metallicity (second and forth panel) distribution for 16 F3D galaxies that have a sufficient number of GC candidates in the ACSFCS catalogue from \cite{Jordan2015} to create a well sampled distribution. MDFs were inferred from the colour distributions using Eq. \ref{eq:piecewise} with the best-fit parameters from our default CZR (Table \ref{tab:colour_met_results}). The histograms are coloured based on the stellar mass of the host using three mass bins: log$(M_\text{Host}/M_\sun) < 10$ (pink), 10 $\leq$ log$(M_\text{Host}/M_\sun) < 11$ (blue), and log$(M_\text{Host}/M_\sun) \geq 11$ (purple). The red histograms show the GCs in our sample. In the top left corner of each panel, the distributions are classified as bimodal ('BI') or unimodal ('UNI') based on the \textsc{GMM} test described in the text.}
\label{fig:color_metal_distribution}
\end{figure*}

\subsection{Dependence of the CZR on host mass}
\label{sect:mass_dependence}
\cite{Usher2015} suggested that different galaxies can show variations in the CZR as a result of different assembly histories. In particular, the mass of the host galaxy might have an influence on the CZR, as was also suggested for M87 \citep{Villaume2019}. With the F3D sample, we can test this to some extent as seen in Fig. \ref{fig:mass_comp}. In this figure, we binned the total GC sample by the stellar mass of the host using three bins. 

The lowest mass bin (log($M_\ast/M_\sun$) $<$ 10) populates the blue, metal-poor end of the CZR. Fitting only these GCs results in a linear CZR with a steep slope similar to that we found at the blue end using the full sample. The linear fit also suggests a lower mean metallicity by $\sim$ 0.15 dex, on the order of the random uncertainties. This offset might reflect the generally younger ages of GCs in low-mass galaxies, but a larger sample would be required to test whether this offset is real.

The intermediate mass bin (10 $\leq$ log($M_\ast/M_\sun$) $<$ 11) contains the largest number of GCs and samples the full range of metallicities and colours. The bent shape of the CZR is visible in this mass bin and does not differ significantly from the one using the full sample. 

The highest mass bin (log($M_\ast/M_\sun$) $>$ 11) contains the GCs of the central galaxy FCC\,213 (NGC\,1399), FCC\,219 and FCC\,167. In this mass range, also a broad range of metallicities and colours is found, but the bluest, most metal-poor GCs are lacking ($g-z < 0.9$ and [M/H] $< -2.0$ dex). Nonetheless, using only these GCs still results in a non-linear CZR that is very similar to that of the full sample.

Although the least massive galaxies in our sample would possible result in a linear CZR simply due to the lack of red, metal-rich GCs, we cannot find indications of different CZR shapes as a function of stellar mass. GCs of the intermediate and high mass galaxies lie on the same track. Nonetheless, the deviations we found with some literature CZRs could lie in the different galaxy masses that are probed. M87 studied by \cite{Villaume2019} is a giant ETG several times more massive than even FCC\,213 (e.g. \citealt{Wu2006, Forte2012}), so we cannot directly compare our GCs to those of M87. 
Also the CZR of \cite{Usher2012} is based on five galaxies with $M_\ast > 10^{10.5} M_\sun$ \citep{Forbes2017a, Tamm2012}. The CZR of \cite{Harris2017} is based on Centaurus A which has a stellar mass $M_\ast \sim 10^{11} M_\sun$ \citep{Woodley2006, Woodley2007, Woodley2010}. 

Alternatively, the different CZRs could reflect different environments.  Many of the galaxies studied in the literature are the most massive galaxy of their group or cluster and thus might have unique assembly histories. In our sample, FCC\,213 only constitutes a fraction of GCs and we could also include the GCs of low-mass galaxies with stellar masses $< 10^{10}M_\sun$, a regime where the CZR of GCs is not yet explored.
Using photometry of GCs in the core region of the Virgo in comparison to the MW, \cite{Powalka2016} found indications that the environment has an influence on colour-colour relations that in part are caused by chemical abundance variations. Therefore, they argue that relations derived in one environment might not be applicable to other environments. The brightest GCs in our sample, however, appear to follow the same [Fe/H] - [M/H] relation as the stars in the MW, although a larger sample would be required to further determine the abundance pattern of GCs in relation to the environment.

\subsection{Globular cluster metallicity distributions}
\label{sect:MDFs}
The non-linear CZR of the F3D GCs has consequences for the MDFs of these galaxies. We can use our empirical relation to translate the extensive photometric GC catalogues from \cite{Jordan2015} to metallicity distributions and hence study the effect of this CZR in more detail. As examples, we picked 16 galaxies that were part of F3D and the ACSFCS to translate their ACSFCS GC colour distributions into metallicity distributions using Eq. \ref{eq:piecewise} with the best-fitting parameters from our CZR. We selected all GC candidates with a probability of being a GC ($p$GC) greater than 50\% (see \cite{Jordan2015} for details). This yields predictions for the metallicity distributions that would satisfy our CZR, shown in Fig. \ref{fig:color_metal_distribution}. The galaxies in this figure are ordered by increasing stellar mass. For comparison, we also show the confirmed GCs in our sample. We note again that our sample is deficient in the bluest GCs ($g - z < 0.8$ mag), possibly also because F3D covers the central parts of galaxies, while the relative number of blue GCs typically increases with galactocentric radius (e.g. \citealt{Faifer2011}). Additionally, because the spatial coverage of the ACSFCS catalogue is limited, the ACSFCS colour distributions shown here might be deficient in blue GCs compared to the total GC distribution. However, these blue GCs are likely to be metal-poor and thus should not affect the bent shape of the CZR.

We quantified the shapes of the colour and metallicity distributions using the Gaussian mixture modelling (\textsc{GMM}) algorithm of \cite{Muratov2010}. This algorithm is a improved version of the \textsc{KMM} code \citep{Ashman1992} and can be used to test whether a distribution is uni- or bimodal. \textsc{GMM} determines the best-fitting parameters of a unimodal and bimodal distribution and uses a bootstrap method to determine whether the bimodal solution is preferred. Following the suggestions of \cite{Muratov2010}, we consider a distribution to be bimodal if the distribution has a negative kurtosis, the relative distance between the two peaks is $D > 2$, and the bimodal solution is preferred with a probability $p > 0.9$. In Fig. \ref{fig:color_metal_distribution}, we noted the bimodal and unimodal distributions with 'BI' and 'UNI', respectively. In the \textsc{GMM} test, we assumed equal-width modes (homosedastic case) and this choice can influence the result \citep{Beasley2018}. 

The \textsc{GMM} test shows that the lower mass galaxies tend to have unimodal colour distributions with a dominant peak at $\sim 0.9$ mag, while the high-mass galaxies have bimodal colour distributions. The relative number of red GCs increases with galaxy mass. The MDFs, however, show a more diverse behaviour because of the non-linear shape of our CZR that smears out blue peaks to broad metal-poor distributions. At low galaxy masses, our CZR at blue colours translates the blue, unimodal colour distributions to broader unimodal MDFs with a peak at low GC metallicities. At intermediate masses (e.g. FCC\,276, FCC\,147 and FCC\,083, log($M_\ast/M_\sun) \sim 10.5$), bimodal colour distributions with roughly equal numbers of red and blue GCs lead to bimodal MDFs with a broader metal-poor peak. FCC\,170 is an outlier in this, and despite its high stellar mass log($M_\ast/M_\sun) \sim 10.4$ \citep{Iodice2019}, shows a unimodal blue distribution with a relatively low number of GCs in total. 

At the highest galaxy masses (log($M_\ast/M_\sun) > 11$, FCC\,167, FCC\,219, and FCC\,213), the minor blue peak is smeared out to a tail of metal-poor GCs, resulting in unimodal MDFs with a dominant peak at high GC metallicities.
This comparison shows that even with this non-linear CZR, not only unimodal MDFs are found. Instead, a diversity of MDFs is expected from our CZR and their modality appears to depend on the host galaxy.

\subsection{Implications for galaxy assembly}
\label{sect:implications}
Most galaxy formation theories explain GC colour and consequently metallicity bimodality by the existence of two distinct populations with different mean metallicities that are connected to different formation places. The bimodality has been linked to a two-stage formation scenario for massive galaxies (e.g. \citealt{Ashman1992, Forbes1997, Cote1998, Beasley2002, Brodie2006, LeeJang2016, Beasley2018}) and is also expected in the hierarchical merger scheme of galaxy formation (e.g. \citealt{Muratov2010, Tonini2013, Li2014, Choksi2018, Kruijssen2019}). It is assumed that the red, metal-rich GCs either form in-situ in massive halos around the peak of star formation or during major mergers of gas-rich galaxies together with the bulk of in-situ stars. They share the high metallicity of the stars because both are set by the local mass-metallicity relation (e.g. \citealt{Shapiro2010}). In contrast, the metal-poor GCs form in smaller haloes from metal-poor gas and are accreted to the main galaxy in a series of hierarchical mergers (see also \citealt{ForbesRemus2018}).

As consequence of the steep slope of our CZR at blue colours, it predicts unimodal MDFs with a broad metal-poor component for galaxies with low mass and a low fraction of red GCs. In contrast, truly bimodal MDFs are expected for intermediate massive galaxies that have roughly a similar number of red and blue GCs, while at the highest galaxy masses, unimodal MDFs with a peak at high ($\sim$ solar) metallicities are expected. 
In context of hierarchical assembly scenarios, this CZR still allows to conclude that the reddest GCs were formed in-situ and the bluest, most metal-poor GCs were formed in metal-poor dwarfs. This conclusion is also supported, for example, by the often observed different radial profiles of both components (e.g. \citealt{Harris2009a, Harris2009b, Faifer2011}) and different kinematics (e.g. \mbox{\citealt{Schuberth2010, Strader2011, Pota2013}}). As we showed in the paper I, especially red GCs trace the metallicity of the host galaxy, as would be expected from an in-situ population, while the blue GCs show large metallicity differences.

However, for GCs of intermediate colours, their origin is less clear than a bimodal colour distribution would suggest because they fall in the region of the CZR that shows a steep slope and thus can have a large range of metallicities. This could indicate that those GCs are a mixed population of both in-situ and ex-situ GCs. For example, they could consist of a population of more metal-poor GCs that has formed in-situ very early-on from less enriched gas, or they are the relatively more metal-rich GCs accreted from more massive satellites. The unimodal MDFs of the most massive galaxies could then be an effect of a rich merger history during which the GCs of galaxies with different but mostly high masses were accreted, while the bimodal MDF of lower mass galaxies were created by a larger number of minor mergers (e.g. \citealt{Xu2012, OLeary2020}). Nonetheless, the merger history of individual galaxies can be very diverse as cosmological simulations suggest and thus a model of the merger history would be required to interpret colour and metallicity distributions. 

As an alternative to the two-phase scenarios, \cite{Yoon2006} showed that a strongly non-linear CZR can create a bimodal colour distribution from a unimodal MDF (see also \citealt{Yoon2011a, Yoon2011b, Kim2013, Chung2016}) without invoking the presence of two distinct populations. Instead, they proposed theoretical non-linear CZRs based on detailed stellar population modelling. Recently, \citet{Lee2019} modelled the colour distributions of a large number of galaxies in the Fornax and Virgo clusters and found that most of the GC system colour bimodality can be explained by unimodal MDFs and a non-linear CZR. They attribute the observed diversity in colour distributions to the mean metallicity of the GC system, where more massive galaxies have a more metal-rich GC system. Our non-linear CZR indeed finds unimodal, metal-poor MDFs for the least massive and unimodal, metal-rich MDFs for the most massive galaxies, in accordance to this picture. However, for intermediate mass galaxies, we still find bimodal MDFs and in the high mass galaxies, we still observe a tail of more metal-poor GCs. 
Although it is possible that this tail consists of only GCs that were formed in-situ under different conditions, the bimodal MDFs in less massive galaxies rather supports the idea of distinct populations, although with less strict metallicity differences than the colour distributions might suggest. This in agreement with the results from recent hydrodynamical simulations that have shown that a one-to-one relation between metallicity and in-situ or accreted population is not given \citep{ForbesRemus2018}.

\section{Conclusions}
\label{sect:conclusions}
We have studied the colour-metallicity relation (CZR) from a sample of 187 GCs of 23 galaxies in the Fornax cluster that were observed as part of the F3D project. These galaxies cover a range in stellar masses between 10$^{8}$ and 10$^{11} M_\sun$.
Our main results are as follows:

\begin{itemize}
\item{We derived metallicities with full spectral fitting and compared them to photometry mainly from the ACSFCS ($g - z$ colours, \citealt{Jordan2015}). The resulting CZR is non-linear. It is shallow at red colours and significantly steepens at bluer colours. The relation can be described by a quadratic function or a piecewise linear function with a breakpoint at ($g - z$) $\sim$ 1.1 mag. A linear relation is not sufficient to describe the shape of the CZR.}


\item{Although our default approach assumes a GC age $\geq$ 8 Gyr, we tested this assumption by also fitting the GC ages. This shows that the metallicities and the CZR are insensitive to the age prior, and the best-fitting ages are old \mbox{($\geq$ 8 Gyr)} with very few exceptions. We only found a weak age-metallicity relation that appears to be mostly driven by the mass of the host because the low mass galaxies in our sample tend to have younger, more metal-poor GCs.}

\item{Using a small sub-sample of the very brightest GCs, we derived [$\alpha$/Fe] abundances and found a negative correlation with metallicities. The more metal-poor GCs seem to be more $\alpha$-enhanced.}

\item{We derived the MZR and found a weak correlation between GC mass and metallicity, in agreement with previous studies. This finding motivates to also apply the CZR to fainter GCs missing from our sample due to sensitivity and S/N limitations. These fainter GCs should sample the same colour range, but might be slightly more metal-poor due to this MZR. It is unlikely that incompleteness affects the shape of the CZR.}

\item{Our CZR generally agrees with literature CZRs at red colours and high metallicities, while there are larger deviations at bluer colours and lower metallicities. We discuss that this might be an effect of the different galaxy masses probed in different studies. Since our sample also includes usually unexplored low-mass galaxies, we were able to measure the metallicities of a large number of blue GCs. When we binned the sample by host mass, we found the same non-linear CZR even for the most massive galaxies.}

\item{Applying the non-linear CZR to photometric GC colour distributions predicts a diversity of MDFs. The shape of the CZR implies that massive galaxies with relatively small blue GC populations have a unimodal MDF with a peak at high and a tail towards lower metallicities. Galaxies with equal numbers of red and blue GCs can truly have a bimodal metallicity distribution, while low mass galaxies show a unimodal MDF with a metal-poor peak, resulting from the lack of red GCs.}

\item{In the context of galaxy assembly, the MDFs predicted by our CZR support different origins for GCs at the metal-poor and metal-rich end of the distribution. While the most-metal rich GCs are likely to have formed in-situ in the host galaxy, the most metal-poor GCs were possibly accreted from low-mass dwarf galaxies. However, the shape of the CZR allows a variety of metallicities for GCs with intermediate colours and this could indicate a diverse origin for these GCs. They might be a mixture of more metal-poor GCs reflecting the metal-poor end of the in-situ GC distribution and the relatively more metal-rich GCs accreted from more massive galaxies.}
\end{itemize}

GCs are important tracers of galaxy assembly and to use them to their full capacity, constraining the CZR is a crucial step. In this work, we could derive a non-linear CZR in the Fornax cluster, using galaxies of a variety of different masses that challenges the simplistic division of GCs into in-situ and accreted solely based on their colour. Although studies in different environments and including more low-mass host galaxies are still needed, the CZR shows that modelling individual merger histories is required to interpret colour and metallicity distributions.

\begin{acknowledgements}
We thank the anonymous referee for comments that have helped to polish this manuscript. GvdV acknowledges funding from the European Research Council (ERC) under the European Union's Horizon 2020 research and innovation programme under grant agreement No 724857 (Consolidator Grant ArcheoDyn). RMcD is the recipient of an Australian Research Council Future Fellowship (project number FT150100333). J. F-B  acknowledges support through the RAVET project by the grant AYA2016-77237-C3-1- P from the Spanish Ministry of Science, Innovation and Universities (MCIU) and through the IAC project TRACES which is partially supported through the state budget and the regional budget of the Consejer\'{i}a de Econom\'{i}a, Industria, Comercio y Conocimiento of the Canary Islands Autonomous Community. IMN acknowledges support from the AYA2016-77237-C3-1-P grant from the Spanish Ministry of Economy and Competitiveness (MINECO) and from the Marie Sk\l odowska-Curie Individual {\it SPanD} Fellowship 702607. EMC is supported by MIUR grant PRIN 2017 20173ML3WW\_001 and by Padua University grants DOR1715817/17, DOR1885254/18, and DOR1935272/19. This research made use of Astropy,\footnote{\url{http://www.astropy.org}} a community-developed core Python package for Astronomy \citep{astropy:2013, astropy:2018}.
\end{acknowledgements}

\bibliographystyle{aa} 
\bibliography{References}

\appendix

\section{Different metallicity measurements}
\label{app:SSP_models}

\begin{table}
\centering
\caption{Overview of different approaches to determine metallicities from the GCs. Method A is the default approach, as also used in paper I.} 
\label{tab:approaches}
\begin{tabular}{c l l} \hline
& Method name & Description \\ \hline \hline
A &  E-MILES & baseFe, age $\geq$ 8 Gyr \\
B & scaled solar MILES & [$\alpha$/Fe] = 0 dex, age $\geq$ 8 Gyr \\
C & $\alpha$-enhanced MILES & [$\alpha$/Fe] = 0.4 dex, age $\geq$ 8 Gyr \\
D & full E-MILES & baseFe, no age constraint\\
E & Line-strength indices & age $>$ 10 Gyr \\ \hline
\end{tabular}
\end{table}

Our default approach to measure GC metallicities used the E-MILES SSP models with an age constraint $\geq$ 8 Gyr. In the following, we present the CZR using different approaches to fit metallicities based on a smaller sub-sample of 135 GCs with S/N $>$ 10 \AA$^{-1}$. Besides E-MILES models with and without age constraint, we also used the scaled solar MILES and the $\alpha$-enhanced MILES models. They have smaller wavelength range and have [$\alpha$/Fe] = 0 dex (scaled solar) and 0.4 dex ($\alpha$-enhanced) at all metallicities, respectively. 

In addition to full spectral fitting, we also determined metallicities of 135 GCs with S/N $>$ 10 \AA$^{-1}$ from our sample using line-strength indices following the method described in \cite{Iodice2019} and \cite{F3D_Survey}. To avoid contamination from sky residuals, a restricted wavelength region between 4800 and 5500 \AA\, was used. The line-strengths of H$\beta$, Fe5015, Mg$b$, Fe5720, and Fe5335 were determined in the LIS system \citep{Vazdekis2010, Vazdekis2015} and were compared to the predictions from the MILES models \citep{Vazdekis2012}.
The best-fitting values were determined using a Markov-Chain-Monte-Carlo algorithm \citep{MartinNavarro2018}.

\begin{figure*}
\centering
\includegraphics[width=0.99\textwidth]{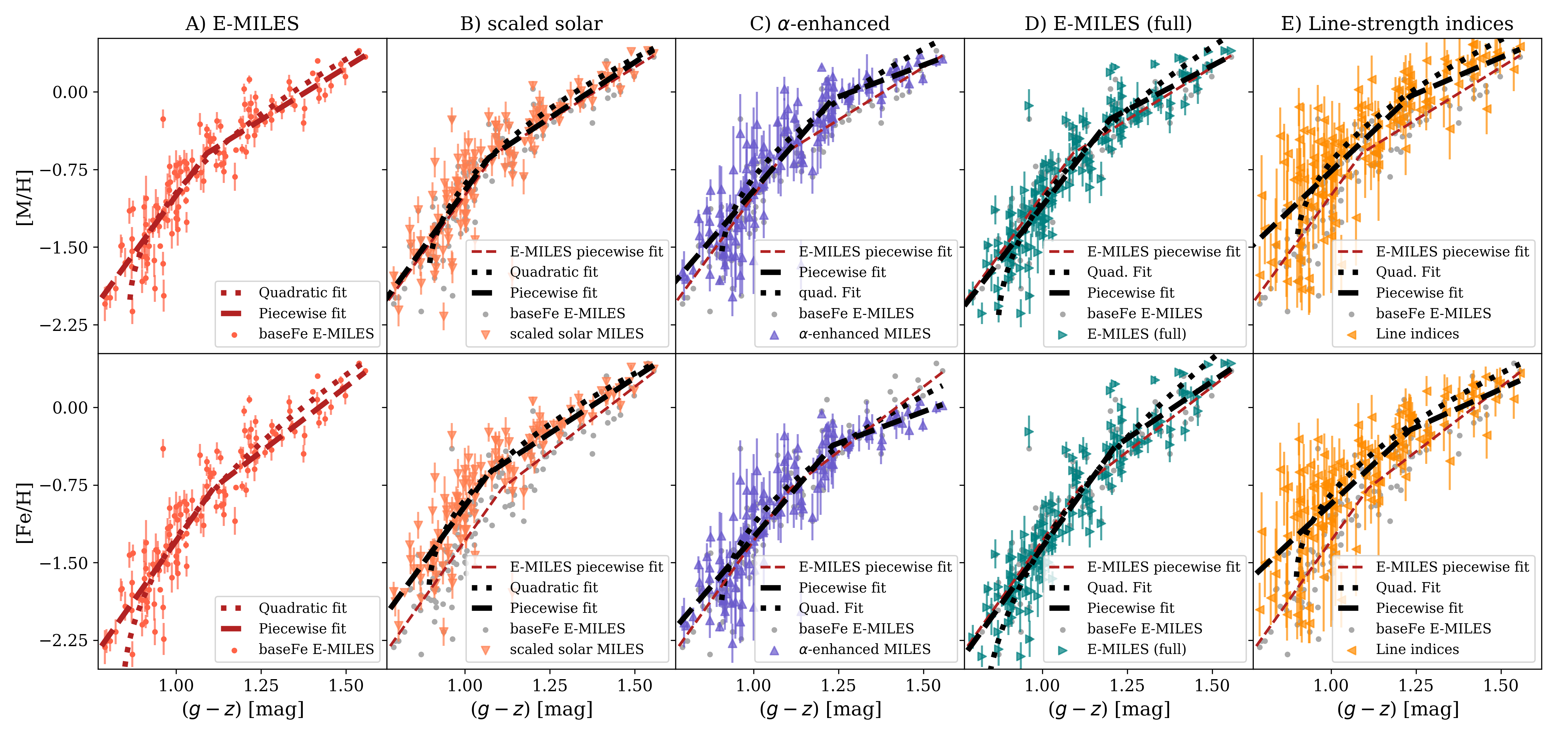}
\caption{CZR for different metallicity measurement approaches (listed in Table \ref{tab:approaches}) based on total metallicities (\textit{top}) and iron metallicities (\textit{bottom}). From left to right: CZR using the E-MILES library with age constraint $\geq$ 8 Gyr (method A), scaled-solar MILES models (method B), $\alpha$-enhanced MILES models (method C), E-MILES models without age constraint (method D), and line-strength indices (method E). The grey symbols show the points from method A as reference, and the lines show the respective fits. The best-fitting parameters are found in Table \ref{tab:colour_met_results}. }
\label{fig:colour_met_relation_comp}
\end{figure*}

Table \ref{tab:approaches} lists the different approaches and we show the resulting CZRs in Fig. \ref{fig:colour_met_relation_comp} for both total and iron metallicities using \mbox{[M/H] = [Fe/H] + 0.75 [$\alpha$/Fe]} as a conversion for the MILES models and the line-strength metallicities and the conversion derived in Sect. \ref{sect:iron_and_total_metallicities} for the E-MILES models. The default values derived with the E-MILES models (method A) are shown as the grey dots. In each case, we fitted the relation both with a quadratic equation (Eq. \ref{eq:quad}) and a piecewise linear curve (Eq. \ref{eq:piecewise}).
The best-fitting parameters from least-square fits to the respective CZRs are reported in Table \ref{tab:colour_met_results}.


Irrespectively of the chosen SSP models, we always found a steep slope of the colour-metallicity relation at low metallicities when using full spectral fitting with \textsc{pPXF}. Using line-strength indices (method E) to measure metallicities results in a significantly larger scatter and larger errorbars, possibly due to the limited wavelength range that is used, but we observed the same non-linear trend in the CZR. 

\begin{table*}
\centering
\begin{threeparttable}
\caption{Colour metallicity relation fit parameters when using total metallicities [M/H] or iron metallicities [Fe/H].}
\label{tab:colour_met_results}

\begin{tabular}{l | c c c | c c c c c}\hline
Method & $a$ & $b$ & $c$ & $m_1$ & $b_1$ & $m_2$ & $b_2$ & $x_0$  \\ \hline
$[$M/H$]$ \\ \hline \hline
A)   & 1.34 $\pm$ 0.01 & 0.46 $\pm$ 0.02 & 0.11 $\pm$ 0.01 & 4.51 $\pm$ 0.32 & $-$5.51 $\pm$ 0.36 & 2.03 $\pm$ 0.20 & $-$2.81 $\pm$ 0.36 & 1.09 $\pm$ 0.03\\
B)   & 1.33 $\pm$ 0.01 & 0.49 $\pm$ 0.03 & 0.14 $\pm$ 0.01 & 4.51 $\pm$ 0.49 & $-$5.47 $\pm$ 0.54 & 2.15 $\pm$ 0.26 & $-$2.94 $\pm$ 0.54 & 1.07 $\pm$ 0.04 \\
C)   &  1.32 $\pm$ 0.01 & 0.48 $\pm$ 0.03 & 0.13 $\pm$ 0.01 & 4.58 $\pm$ 0.49 & -5.53 $\pm$ 0.55 & 2.16 $\pm$ 0.26 & $-$2.94 $\pm$ 0.54 & 1.07 $\pm$ 0.04\\
D)  & 1.31 $\pm$ 0.01 & 0.39 $\pm$ 0.02 & 0.08 $\pm$ 0.01 & 4.21 $\pm$ 0.27 & $-$5.31 $\pm$ 0.35 & 1.72 $\pm$ 0.39 & $-$2.32 $\pm$ 0.35 & 1.20 $\pm$ 0.04 \\ 
E)   & 1.24 $\pm$ 0.01 & 0.45 $\pm$ 0.03 & 0.15 $\pm$ 0.02 & 3.11 $\pm$ 0.29 & $-$3.88 $\pm$ 0.45 & 1.41 $\pm$ 0.71 & $-$1.77 $\pm$ 0.44 & 1.23 $\pm$ 0.08  \\ 
\hline
$[$Fe/H$]$ \\ \hline \hline
A) & 1.37 $\pm$ 0.01 & 0.38 $\pm$ 0.02 & 0.07 $\pm$ 0.01 & 4.63 $\pm$ 0.31 & $-$5.90 $\pm$ 0.38 & 2.51 $\pm$ 0.24 & $-$3.55 $\pm$ 0.37 & 1.11 $\pm$ 0.03\\
B) & 1.33 $\pm$ 0.01 & 0.49 $\pm$ 0.03 & 0.14 $\pm$ 0.01 & 4.51 $\pm$ 0.49 & $-$5.47 $\pm$ 0.54 & 2.15 $\pm$ 0.26 & $-$2.94 $\pm$ 0.54 & 1.07 $\pm$ 0.04 \\
C) & 1.47 $\pm$ 0.02 & 0.55 $\pm$ 0.03 & 0.13 $\pm$ 0.01 & 3.69 $\pm$ 0.26 & $-$5.00 $\pm$ 0.36 & 1.64 $\pm$ 0.47 & $-$2.51 $\pm$ 0.35 & 1.21 $\pm$ 0.05 \\
D) & 1.34 $\pm$ 0.01 & 0.32 $\pm$ 0.02 & 0.05 $\pm$ 0.01 & 4.55 $\pm$ 0.27 & $-$5.90 $\pm$ 0.38 & 2.21 $\pm$ 0.47 & $-$3.06 $\pm$ 0.37 & 1.21 $\pm$ 0.04 \\
E) & 1.32 $\pm$ 0.01 & 0.51 $\pm$ 0.03 & 0.15 $\pm$ 0.02 & 3.07 $\pm$ 0.29 & $-$4.00 $\pm$ 0.46 & 1.48 $\pm$ 0.71 & $-$2.04 $\pm$ 0.45 & 1.23 $\pm$ 0.09 \\
\end{tabular}
\begin{tablenotes}
      \item The parameters $a$, $b$, and $c$ refer to least-square fits with a quadratic equation (Eq. \ref{eq:quad}), the $m_1$, $b_1$, $m_2$, $b_2$, and $x_0$ to the piecewise linear fit (Eq. \ref{eq:piecewise}). The different methods are described in Table \ref{tab:approaches}.
    \end{tablenotes}
\end{threeparttable}
\end{table*}

\end{document}